\documentclass[a4paper,12pt]{article}
\pdfoutput=1
\usepackage[a4paper,left=0.99in, right=0.99in,top=1.2in, bottom=1.2in]{geometry}

\usepackage{amsmath}
\usepackage{hyperref}
\usepackage{color}
\usepackage{cite}
\usepackage{graphicx}

\newcommand{\nn}{\nonumber}

\newcommand{\beq}{\begin{equation}}
\newcommand{\eeq}{\end{equation}}
\newcommand{\bea}{\begin{eqnarray}}
\newcommand{\eea}{\end{eqnarray}}
\newcommand{\beqa}{\begin{eqnarray}}
\newcommand{\eeqa}{\end{eqnarray}}

\definecolor{red}{rgb}{1,0,0}

\def\be{\begin{equation}}
\def\ee{\end{equation}}
\numberwithin{equation}{section}

\title{
\vskip 1cm
Four-jet production in single- and double-parton scattering within high-energy factorization
\vskip 1cm
}
\author{
Krzysztof Kutak$^1$, Rafal Maciula$^1$,
Mirko Serino$^1$, \\ Antoni Szczurek$^{1,2}$ and Andreas van Hameren$^1$ \\ \\
$^1$ {\small\it The H.\ Niewodnicza\'nski Institute of Nuclear Physics, Polish Academy of Sciences,}\\ {\small\it Radzikowskiego 152, 31-342 Krak\'ow, Poland}\\\\
$^2$ {\small\it 
University of Rzesz\'ow, PL-35-959 Rzesz\'ow, Poland}\\
}

\date{}

\begin{document}
\maketitle

\thispagestyle{empty}

\vspace{-28em}
\begin{flushright}
 IFJPAN-IV-2016-2 \\
\end{flushright}
\vspace{14em}

\vspace{10em}
\begin{abstract}

We perform a first study of 4-jet production in a complete high-energy
factorization (HEF) framework. We include and discuss contributions
from both single-parton scattering (SPS) and double-parton scattering
(DPS). The calculations are performed for kinematical situations
relevant for two experimental measurements (ATLAS and CMS) at the LHC.
We compare our results to those reported by the ATLAS and CMS
collaborations for different sets of kinematical cuts. The results of
the HEF approach are compared with their counterparts for collinear factorization.
For symmetric cuts the DPS HEF result is
considerably smaller than the one obtained with collinear
factorization. The mechanism leading to this difference is of
kinematical nature. We conclude that an analysis of inclusive 4-jet
production with asymmetric $p_T$-cuts below 50 GeV would be useful to
enhance the DPS contribution relative to the SPS contribution. In
contrast to the collinear approach, the HEF approach nicely describes
the distribution of the $\Delta S$ variable, which involves all four
jets and their angular correlations.

\end{abstract}

\section{Introduction}

So far, complete $(n\geq4)$-jet production via single-parton scattering (SPS) was
discussed only within collinear factorization. Results up to next-to-leading (NLO)
precision can be found in \cite{Bern:2011ep,Badger:2012pf}.
Here we wish to discuss for the first time production of four jets within high-energy ($k_T$-)factorization (HEF)
approach with 2 $\to$ 4 subprocesses with two
off-shell partons. Recently three of us have discussed another reaction with
2 $\to$ 4  ($g g \to c \bar c c \bar c$) subprocess in the framework
of the HEF \cite{vanHameren:2015wva}.
For the four-jet production the number of subprocesses is much higher. 

Double-parton scattering (DPS) was claimed to have been observed for 
the first time at the Tevatron \cite{Abe:1993rv}.
In the LHC era, with much higher collision energies available, 
the field has received a new impulse and 
several experimental and theoretical studies address the problem
of pinning down DPS effects (for review see \cite{Diehl:2011yj,Bansal:2014paa}).

Even just from purely theoretical point of view, the problem is 
quite subtle.
As for the non perturbative side, it is in principle necessary, when
considering a double-parton scattering, to take into account 
the correlations between the two partons coming from the same protons 
and involved in the scattering processes. 
Such an information should be encoded in a set of double parton
distribution functions (DPDFs), generalising usual parton distribution functions (PDFs).
A benchmarking work on DPDFs was made in
Ref.~\cite{Gaunt:2009re}, where a proper generalisation of the DGLAP
evolution equations to DPDFs was provided. 
Building explicit initial conditions for the evolution equations
is challenging. Some successful attempts are becoming 
to appear only recently 
\cite{Golec-Biernat:2015aza,Rinaldi:2014ddl,Broniowski:2016trx}.
In the meanwhile, phenomenological and experimental studies of double-parton scattering
rely on factorized Ansatz for the DPDFs, which amount
to neglecting longitudinal momentum correlations
between partons and treating transversal ones by introducing an
effective cross section, $\sigma_{eff}$. 
The latter quantity is usually extracted from experimental data.
In the present approach we will use the factorized Ansatz and
concentrate on the difference between leading-order collinear and
high-energy-factorization results. The latter includes effectively
higher-order corrections. 
For most of high-energy reactions the single-parton scattering dominates
over the double-parton scattering. The extraordinary example
is double production of $c \bar c$ pairs \cite{Luszczak:2011zp,Maciula:2013kd}.
For four-jet production, disentangling the ordinary SPS
contributions from the DPS corrections can be quite challenging for
several reasons: first of all, it is
necessary to define sufficiently sensitive, process-dependent obervables, 
w.r.t. which the DPS differential cross section manifestly dominates at
least in some corners of phase space \cite{Berger:2009cm,Maciula:2015vza}.
Nevertheless, even once this is done, one has to be careful about the
kinematical regime employed in comparing experimental data to
theoretical prediction: in fact, the generally decreasing 
behaviour of PDFs for large momentum fractions \cite{Martin:2009iq} is well known, 
particularly for gluons, and gluon-initiated processes account for a very large part of the cross section;
this implies that, for very energetic final states (characterized by large transverse
momenta), it is really unlikely to get contributions from DPS.
This is confirmed very well experimentally by the data released by the
ATLAS Collaboration for both the 7 and 8 TeV runs \cite{Aad:2011tqa,Aad:2015nda}. 
This problem is of course slightly tamed by providing high center of
mass energy in hadron-hadron scattering, as moderately
low values of $x$ should be enough to guarantee observing DPS in a
kinematic regime in which perturbative QCD, possibly supplemented 
by parton showering, still works reasonably well. 

In this paper we propose to assess the predictions of HEF for double-parton scattering at the LHC in a leading-order (LO) framework.
HEF is an approach introduced in the early 90's in the context of 
heavy-flavour production, in order to take into account the effect of the colliding parton transverse momentum, 
which is neglected in the collinear approach \cite{Collins:1991ty,Catani:1990eg,Catani:1994sq}. 
This implies using off-shell partons, for which the construction of
gauge-invariant scattering amplitudes is not straightforward. However,
recent improvements in the understanding of scattering amplitudes
have allowed to formulate efficient analytical and numerical algorithms
for the computation of such objects
\cite{vanHameren:2013csa,vanHameren:2012if,Kotko:2014aba,vanHameren:2014iua,vanHameren:2015bba,
Cruz-Santiago:2015nxa,Cruz-Santiago:2015dla,Kotko:2016qxv}.

With such a machinery, we expand the analysis
presented in Ref.~\cite{Maciula:2015vza} and assess
the differences between the pure collinear approach and the high-energy
factorization (HEF) called also $k_T$-factorization framework.
We shall focus on the difference between predictions of HEF
and standard collinear approach for the DPS contribution.

\section{Single-parton scattering production of four jets}%

The collinear factorization formula for the calculation of the inclusive partonic 4-jet cross section at the Born level reads 
\bea
\sigma^B_{4-jets} 
&=& 
\sum_{i,j} \int \frac{dx_1}{x_1}\,\frac{dx_2}{x_2}\, x_1f_i(x_1,\mu_F)\, x_2f_j(x_2,\mu_F) \nn \\
&&
\times \frac{1}{2 \hat{s}} \prod_{l=i}^4 \frac{d^3 k_l}{(2\pi)^3 2 E_l} \Theta_{4-jet}\,  (2\pi)^4\, \delta\left( x_1P_1 + x_2P_2 - \sum_{l=1}^4 k_i \right)\, 
\overline{\left| \mathcal{M}(i,j \rightarrow 4\, \text{part.})  \right|^2} \, . \nn \\
\label{coll_cross}
\eea
Here $x_{1,2}f_i(x_{1,2},\mu_F)$ are the collinear PDFs
for the $i-th$ parton, carrying $x_{1,2}$ momentum fractions of the proton and evaluated at the factorization scale $\mu_F$;
the index $l$ runs over the four partons in the final state, the
partonic center of mass energy squared is $\hat{s} = 2\,x_1 x_2\, P_i \cdot P_j$; 
the function $\Theta_{4-jet}$ takes into account the kinematic cuts applied and 
$\mathcal{M}$ is the partonic on-shell matrix element, which includes symmetrization effects due to identity of particles in the final state.

Switching to HEF, the analogous formula to (\ref{coll_cross}) looks as follows:
\bea
\sigma^B_{4-jets} 
&=& 
\sum_{i,j} \int \frac{dx_1}{x_1}\,\frac{dx_2}{x_2}\, d^2 k_{T1} d^2 k_{T2}\,  \mathcal{F}_i(x_1,k_{T1},\mu_F)\, \mathcal{F}_j(x_2,k_{T2},\mu_F) \nn \\
&&
\hspace{-25mm}
\times \frac{1}{2 \hat{s}} \prod_{l=i}^4 \frac{d^3 k_l}{(2\pi)^3 2 E_l} \Theta_{4-jet} \, (2\pi)^4\, \delta\left( x_1P_1 + x_2P_2 + \vec{k}_{T\,1}+ \vec{k}_{T\,2} - \sum_{l=1}^4 k_i \right)\, 
\overline{ \left| \mathcal{M}(i^*,j^* \rightarrow 4\, \text{part.})
  \right|^2 } \, . \nn \\
\label{kt_cross}
\eea
Here $\mathcal{F}_i(x_k,k_{Tk},\mu_F)$ is a transverse momentum
dependent (TMD) distribution function for a given type of parton. 
Similarly as in the collinear factorization case, $x_k$ is the longitudinal
momentum fraction, $\mu_F$ is a factorization scale. 
The new degrees of freedom are introduced via $\vec{k}_{Tk}$, 
which are the parton's transverse momenta, 
i.e. the momenta perpendicular to the collision axis. 
The formula is valid when the $x$'s are not too large and not too small when complications from nonlinearities may eventually arise \cite{Kotko:2015ura}\footnote{ For some processes High Energy Factorization can has been shown valid at NLO accuracy \cite{Chirilli:2011km,Nefedov:2015ara}}. 
The TMD parton densities (for a recent review see \cite{Angeles-Martinez:2015sea}) 
can be defined by introducing operators whose expectation values, roughly speaking, count the number of partons \cite{Collins:2011zzd}. 
In particular, an evolution equation for TMDs known as CCFM, valid both in the low $x$ and large $x$ domain,  
\cite{Marchesini:1994wr,Catani:1989sg} provides a gluon density depending  on $x,\,k_{T},\,\mu$. 
However, for our purposes this is not enough, since we want to have access to moderate values 
of $x$ where the CCFM approach needs refinements \cite{Hautmann:2014uua,Gituliar:2015agu}.
The alternative is to use the Kimber-Martin-Ryskin (KMR) prescription \cite{Kimber:2001sc,Kimber:1999xc} 
in order to obtain a full set of TMD parton densities. 
The basic observation is that the $k_T$ dependence can be generated 
at the very last step of the collinear evolution by performing soft gluon resummation between two scales 
given by $k_T$ and $\mu_F$, where $k_{T}$ is interpreted as the transverse momentum 
of the hardest emitted gluon during the partonic evolution, while $\mu_F$ can be linked to hard scattering scale. 
In practical terms, this procedure boils down to applying the Sudakov form factor onto the PDFs (some details can be found in Appendix \ref{App_TMDs}
\footnote{The TMDs can be obtained by request from krzysztof.kutak@ifj.edu.pl}).

$\mathcal{M}(i^*,j^* \rightarrow 4\, \text{part.})$ is the gauge invariant matrix element for $2\rightarrow 4$ particle scattering with two initial off-shell legs.
In the case of HEF (for recent review see Ref.~\cite{Sapeta:2015gee}), amplitudes with external off-shell legs in QCD have been computed with different approaches:
up to $2 \rightarrow 2$ scattering, they are given for example in \cite{Nefedov:2013ywa} and are enough in order to calculate DPS contributions (see section \ref{MPI}).
In order to move on to higher multiplicities, which are necessary for the SPS analysis of $2 \rightarrow 4$ parton scattering, it is possible to generate this amplitudes analytically 
applying suitably defined Feynman rules \cite{vanHameren:2012if,vanHameren:2013csa}.
Also recursive methods have been developed for this purpose, like generalised BCFW recursion \cite{vanHameren:2014iua,vanHameren:2015bba}
or Wilson lines approaches \cite{Kotko:2014aba,Cruz-Santiago:2015nxa,Kotko:2016qxv}.
By now also a numerical package implementing numerical BCFW recursion is available \cite{Bury:2015dla}. 
In this case, we rely on a numerical approach implemented in AVHLIB\footnote{available for download at https://bitbucket.org/hameren/avhlib} 
which employs Dyson-Schwinger recursion generalized to tree-level amplitudes with off-shell initial-state particles.
Originally proposed in~\cite{Berends:1987me,Caravaglios:1995cd}, this recursive method exists in several explicit implementations with on-shell initial-state particles~\cite{Kanaki:2000ey,Mangano:2002ea,Moretti:2001zz,Gleisberg:2008fv,Kleiss:2010hy}, and has even been extended to one-loop amplitudes~\cite{vanHameren:2009vq,Actis:2012qn}.
AVHLIB and the  Monte Carlo program therein are also used to perform the phase-space integration.
In the collinear case, results were cross-checked by comparing them with the ALPGEN output \cite{Mangano:2002ea}.
We use a running $\alpha_s$ provided with the MSTW2008nlo68cl PDF sets and set both the renormalization and factorization
scales equal to half the transverse energy, which is defined as the sum of the final state transverse momenta, 
$\mu_F=\mu_R= \frac{\hat{H}_T}{2} = \frac{1}{2} \sum_{l=1}^4 k_T^l$\footnote{As customary in the literature, 
we use the $\hat{H}_T$ notation to refer to the energies of the final state partons, 
not jets, despite this is obviously the same thing in a LO analysis},  working in the $n_F = 5$ flavour scheme.

In order to cross-check our numerical tools, we must compare their outputs to results
already available in the literature. For this purpose, we compared LO total cross sections 
for $(n \leq 4)$-jet production to those given by the BlackHat collaboration in Ref.~\cite{Bern:2011ep} 
and cross-checked in Ref.~\cite{Badger:2012pf}. We find excellent agreement, up to phase space integration accuracy.
The cuts used in these calculations were those chosen by the ATLAS
collaboration in the 2011 analysis of multi-jet events \cite{Aad:2011tqa}, 
namely $p_T > 80$ GeV for the leading jet and $p_T > 60$ GeV for subleading
jets, $|\eta| < 2.8$ for the pseudorapidity and jet cone radius parameter $\Delta R>0.4$.
Again we find excellent agreement between the two codes with the LO results reported in the literature, 
up to phase space integration uncertainties.

To be precise, we reproduce the LO predictions for the total inclusive cross sections
\bea
\sigma(\geq 2\, \text{jets}) &=& 958(1)^{+316}_{-221} \, ,
\nn \\
\sigma(\geq 3\, \text{jets}) &=& 93.4(0.1)^{+50.4}_{-30.3} \, ,
\nn \\
\sigma(\geq 4\, \text{jets}) &=& 9.98(0.01)^{+7.40}_{-3.95} \, ,
\eea
where the numbers in brackets stand for the numerical integration uncertainty and the upper and lower errors
are obtained by varying the renormalization scale up and down by a factor of two.

There are 19 different channels contributing to the cross section at the parton-level:
\bea
&& 
g g \rightarrow 4 g \, , 
g g \rightarrow q \bar{q} \, 2g \, , 
q g \rightarrow q \, 3g \, , 
q \bar{q} \rightarrow q \bar{q}\, 2g \, , 
q q \rightarrow q q \, 2g \, , 
q q' \rightarrow q q' \, 2g \, , 
\nn \\
&&
g g\rightarrow q \bar{q} q \bar{q} \, ,
g g \rightarrow q \bar{q} q' \bar{q}' \, ,  
q g \rightarrow q g q \bar{q} \, , 
q g \rightarrow q g q' \bar{q}' \, , 
\nn \\
&&
q \bar{q} \rightarrow 4g\, , 
q \bar{q} \rightarrow q' \bar{q}'\, 2g \, , 
q \bar{q} \rightarrow q \bar{q} q \bar{q} \, , 
q \bar{q} \rightarrow q\bar{q} q' \bar{q}' \, , 
q \bar{q} \rightarrow q' \bar{q}' q' \bar{q}' \, , 
q \bar{q} \rightarrow q' \bar{q}' q'' \bar{q}'' \, , 
\nn \\
&&
q q \rightarrow q q q \bar{q} \, ,  
q q \rightarrow q q q' \bar{q}' \, , 
q q' \rightarrow q q' q \bar{q} \, ,
\nn
\eea
The processes in the first line are the dominant channels,
contributing together to $\sim 93 \% $ of the total cross section. 
This stays true in the HEF framework as well.

\section{Double-parton scattering production of four jets}\label{MPI}

Single-parton scattering contributions are expected to be dominant for high momentum transfer, as it is highly unlikely that
two partons from one proton and two from the other one are energetic enough for two hard scatterings to take place,
as  the behaviour of the PDFs for large $x$ suggests.
However, as the cuts on the transverse momenta of the final state are softened, a window opens
to possibly observe significant double parton scattering effects, as
often stated in the literature on the subject
and recently analysed for 4-jet production in collinear factorization approach in Ref.~\cite{Maciula:2015vza}.
Our goal here is to perform the same analysis in HEF, in order to assess the difference in the predictions
of the two approaches.

First of all, let us recall the formula usually employed for the computation of DPS cross sections, 
adjusting it to the 4-parton final state,
\beq
\frac{d \sigma^{B}_{4-jet,DPS}}{d \xi_1 d \xi_2} = 
\frac{m}{\sigma_{eff}} \sum_{i_1,j_1,k_1,l_1;i_2,j_2,k_2,l_2}
\frac{d \sigma^B(i_1 j_1 \rightarrow k_1 l_1)}{d \xi_1}\, \frac{d \sigma^B(i_2 j_2
\rightarrow k_2 l_2)}{d \xi_2} \, ,
\eeq
where the $\sigma(a b \rightarrow c d)$ cross sections are obtained by
restricting formulas (\ref{coll_cross}) and (\ref{kt_cross}) to a single
channel and the symmetry factor $m$ is $1$ unless the two hard
scatterings are identical, in which case it is $1/2$, so as to avoid
double counting them. Above $\xi_1$ and $\xi_2$ stand for generic
kinematical variables for the first and second scattering, respectively.

The effective cross section $\sigma_{eff}$ can be loosely interpreted 
as a measure of the transverse correlation of the two partons inside 
the hadrons, whereas the possible longitudinal correlations are usually
neglected (for an introduction to this issue, see for example Ref.~\cite{Gaunt:2009re}).
In this paper we use $\sigma_{eff}$ provided by the CDF, D0
collaborations and recently confirmed by the LHCb collaboration
$\sigma_{eff}$ = 15 mb, although the latter value may be questioned
\cite{Maciula:2016wci} when all SPS mechanisms of double charm production are included.

As already mentioned in the introduction there are attempts, in the
literature, to construct DPDFs which
include correlations also between the longitudinal momenta of the two
partons and fullfil sum rules. These models are, however, still rather 
at a preliminary stage. 
So far they are formulated exlusively in the gluon sector \cite{Golec-Biernat:2015aza}
or in the valence quark sector \cite{Rinaldi:2014ddl}.
In addition they are only formulated in a leading order framework
which may be not sufficient for many processes.
Moreover, as it is expected on physical grounds and confirmed by 
all the calculations in the various models proposed so far, 
the longitudinal parton-parton correlations should become far less
important as the energy of the collision is increased, due to the increase in the parton multiplicity.
For instance, the plots in Ref.~\cite{Golec-Biernat:2015aza} show 
that the double gluon distribution obtained with a sum rule approach is
essentially equal to the factorized Ansatz at the scale $Q^2 = 10\, \text{GeV}^2$
down to $x = 10^{-5}$.
Looking forward to further improvements in this field, we choose to
limit ourselves to a more pragmatic approach for the purpose of this paper, 
making the following ansatz for DPDF in the collinear-factorization case:
\beq
D_{1, 2}(x_1,x_2,\mu) = f_1(x_1,\mu)\, f_2(x_2,\mu) \, \theta(1-x_1-x_2) \, , 
\eeq
where $D_{1, 2}(x_1,x_2,\mu)$ is the DPDF and
$f_i(x_i,\mu)$ are the ordinary PDFs and the subscripts $1$ and $2$
simply differentiate between two generic partons in the same proton.
this ansatz can be automatically generalised to the case when parton
transverse momenta are included. 

Coming to DPS contributions, we have to include all the possible $45$ channels which
can be obtained by coupling in all possible distinct ways the $8$ channels for the $2\rightarrow 2$
SPS process, i.e.
\bea
\#1 &=& g g \rightarrow g g  \, , \quad \#5 = q \bar{q} \rightarrow q'\bar{q}'  \nn \; , \\
\#2 &=& g g \rightarrow q \bar{q} \, , \quad \#6 = q \bar{q} \rightarrow g g  \nn \;, \\
\#3 &=& q g \rightarrow q g    \, , \quad \#7 = q q \rightarrow q q \nn \; , \\
\#4 &=& q \bar{q} \rightarrow q \bar{q}   \, , \quad \#8 = q q'\rightarrow q q'   \nn \; .
\eea
We find that the pairs $(1,1)$, $(1,2)$, $(1,3)$, $(1,7)$, $(1,8)$, $(3,3)$ $(3,7)$, $(3,8)$ 
together account for more than $95$ \% of the total cross section for all
the sets of cuts considered in this paper.
%

\subsection{Comparison to the collinear approach and to ATLAS data with hard central kinematic cuts}

In the following, we test the HEF calculation against the collinear case and compare it to the 8 TeV
data recently reported by the ATLAS collaboration \cite{Aad:2015nda}. 
The kinematic cuts are here slightly different with respect to Ref.~\cite{Aad:2011tqa}:
$p_T > 100 $ GeV for the leading jet and $p_T > 64 $ GeV for the first three subleading jets; in addition
$|\eta| < 2.8$ is the pseudorapidity cut and $\Delta R > 0.5$ is
the constraint on the jet cone radius parameter.

\begin{figure}[!h]
\begin{minipage}{0.47\textwidth}
 \centerline{\includegraphics[width=1.0\textwidth]{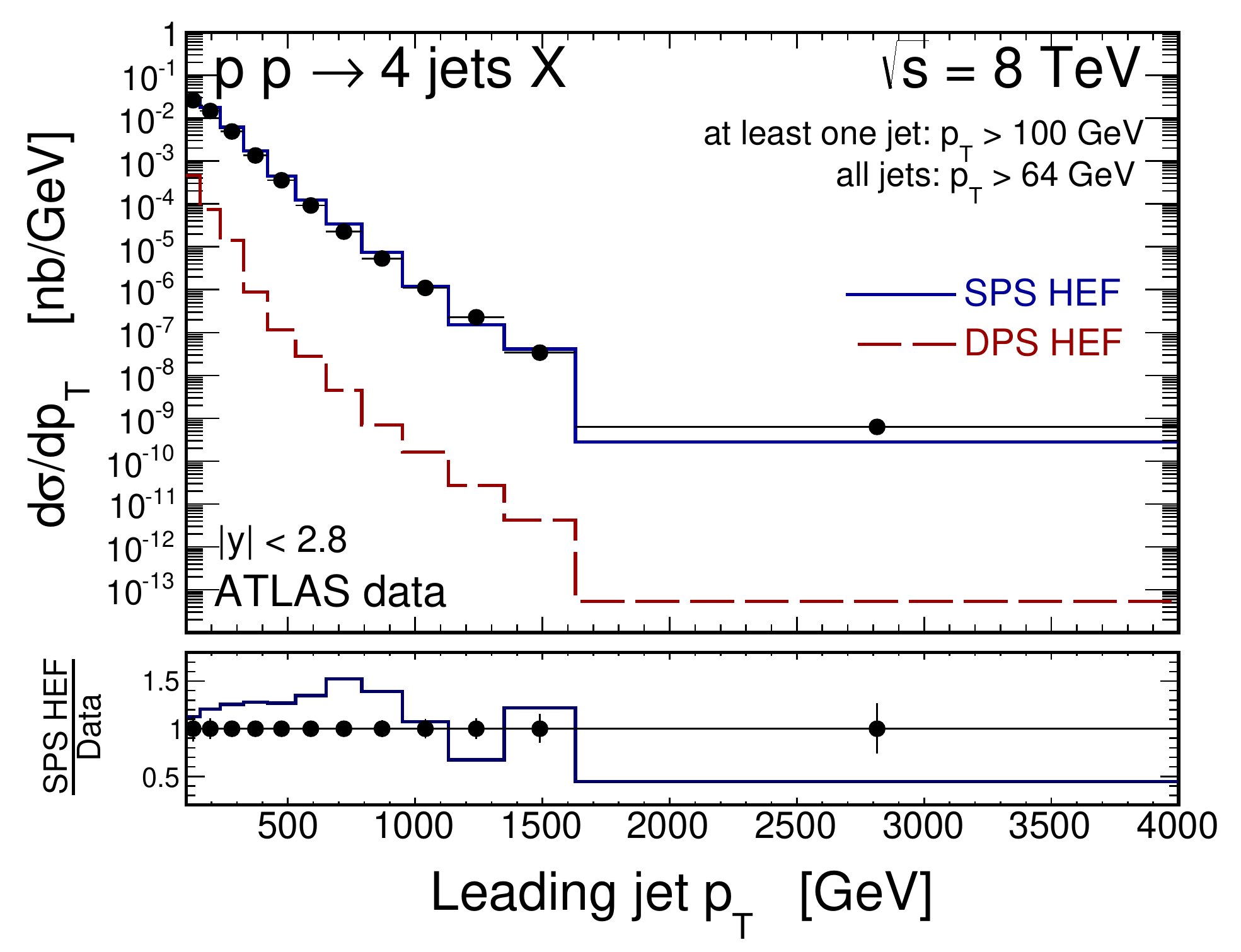}}
\end{minipage}
\hspace{0.5cm}
\begin{minipage}{0.47\textwidth}
 \centerline{\includegraphics[width=1.0\textwidth]{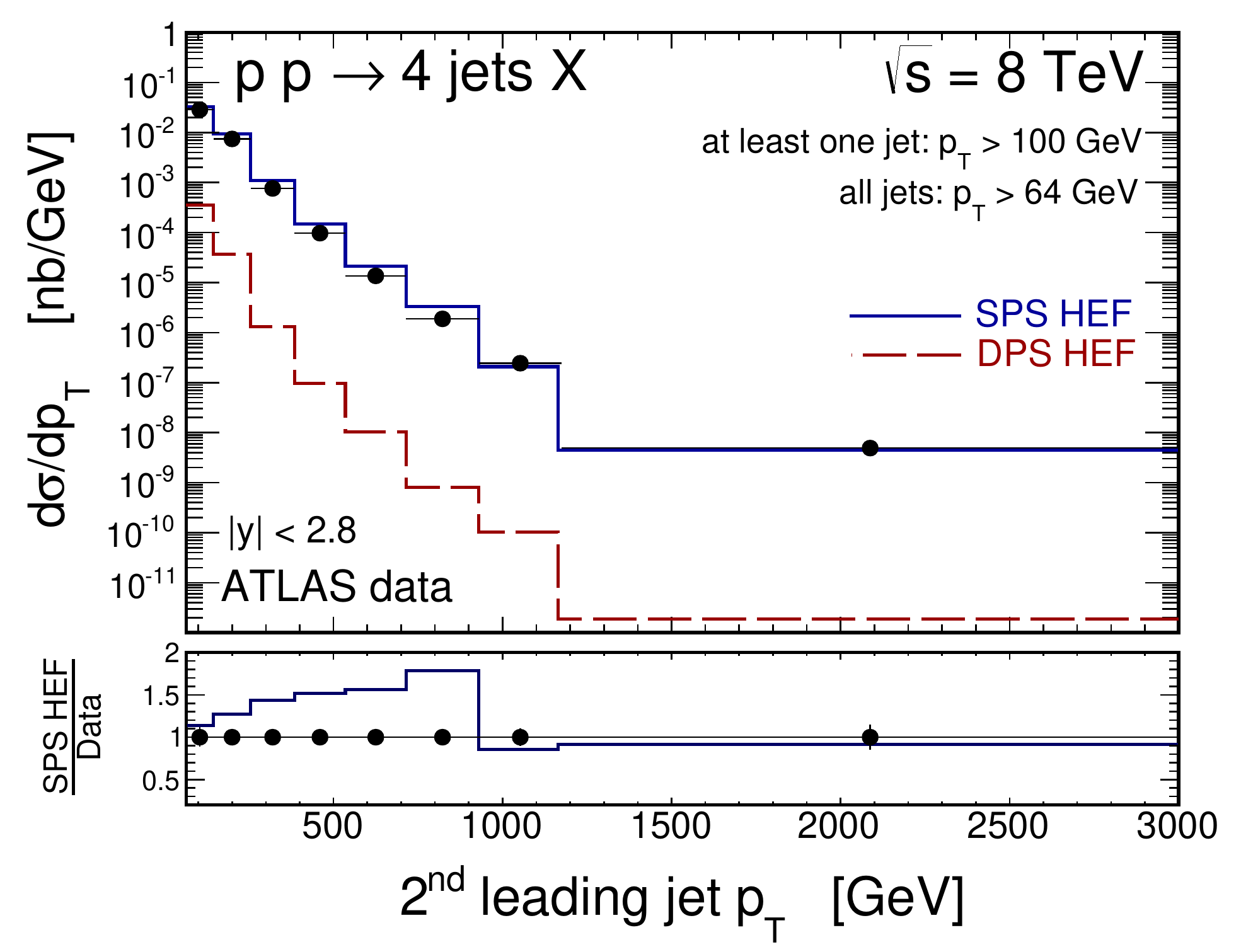}}
\end{minipage}
\caption{
HEF prediction of the differential cross sections w.r.t. the transverse momenta of the first two leading jets
compared to the ATLAS data \cite{Aad:2015nda}. The LO calculation describes the data pretty
well in this hard regime in which MPIs are irrelevant. In addition we show the ratio of the SPS HEF result to the ATLAS data.}
\label{Hard_central_1}
\end{figure}

\begin{figure}[!h]
\begin{minipage}{0.47\textwidth}
 \centerline{\includegraphics[width=1.0\textwidth]{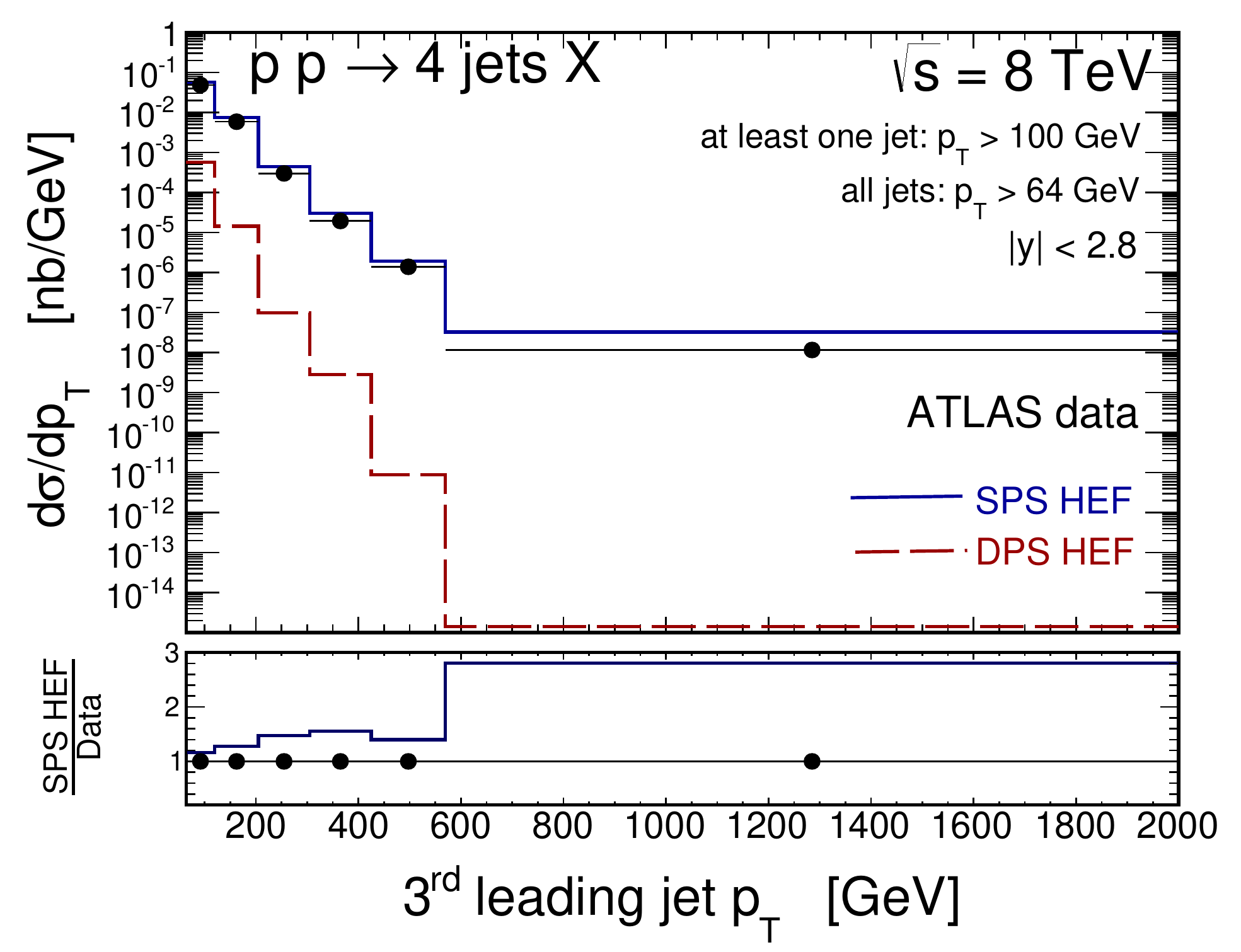}}
\end{minipage}
\hspace{0.5cm}
\begin{minipage}{0.47\textwidth}
 \centerline{\includegraphics[width=1.0\textwidth]{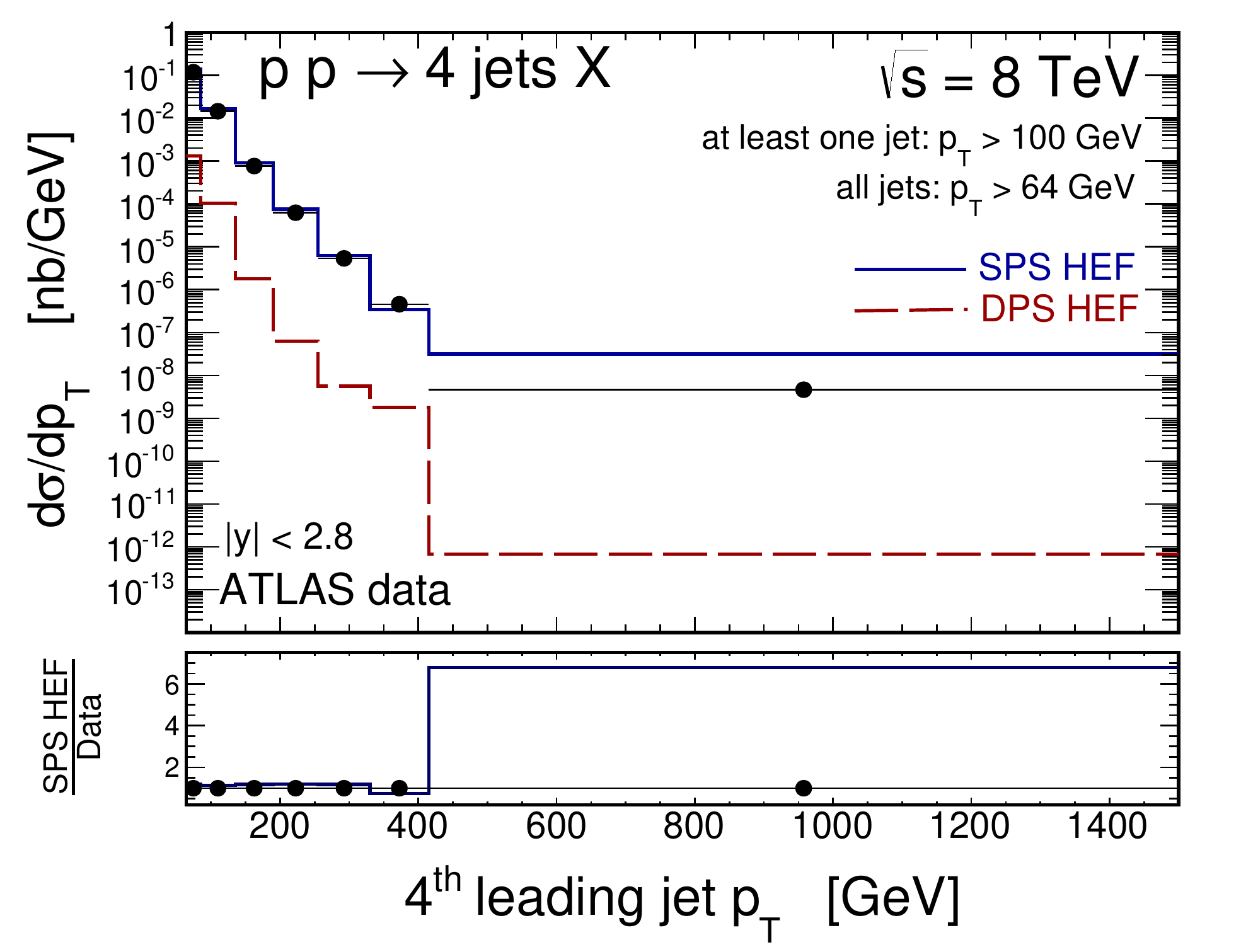}}
\end{minipage}
\caption{
HEF prediction of the differential cross sections w.r.t. the transverse momenta of the 3rd and 4th leading jets
compared to the ATLAS data \cite{Aad:2015nda}. The LO calculation describes the data
pretty well in this hard regime in which MPIs are irrelevant. In addition we show the ratio of the SPS HEF result to the ATLAS data.}
\label{Hard_central_2}
\end{figure}

As for this framework, we employ, together with the newly obtained TMD
PDFs ($5$ quark flavors and gluon) which we call DLC-2016
(Double-Log-Coherence), the running $\alpha_s$ coming with the MSTWnlo200868cl sets.
The results of our computation in HEF is shown in Figs.~\ref{Hard_central_1} and \ref{Hard_central_2},
where it is apparent that the DPS contribution is completely irrelevant, as expected for final states
with high transverse momenta, as it is extremely unlikely that all the four partons in the two couples 
coming from the colliding protons carry enough energy to produce such a hard final state.
A generally good agreement with the ATLAS data can be seen through the transverse momenta
spectra of the four jets, thus showing that the HEF approach works reliably well in this region.

First we show the results of the HEF calculation in Figs.~\ref{Hard_central_1} and \ref{Hard_central_2}.
The prediction is consistent with the ATLAS data for all the $p_T$ spectra. 
%
%

\begin{figure}[h]
\begin{minipage}{0.47\textwidth}
 \centerline{\includegraphics[width=1.0\textwidth]{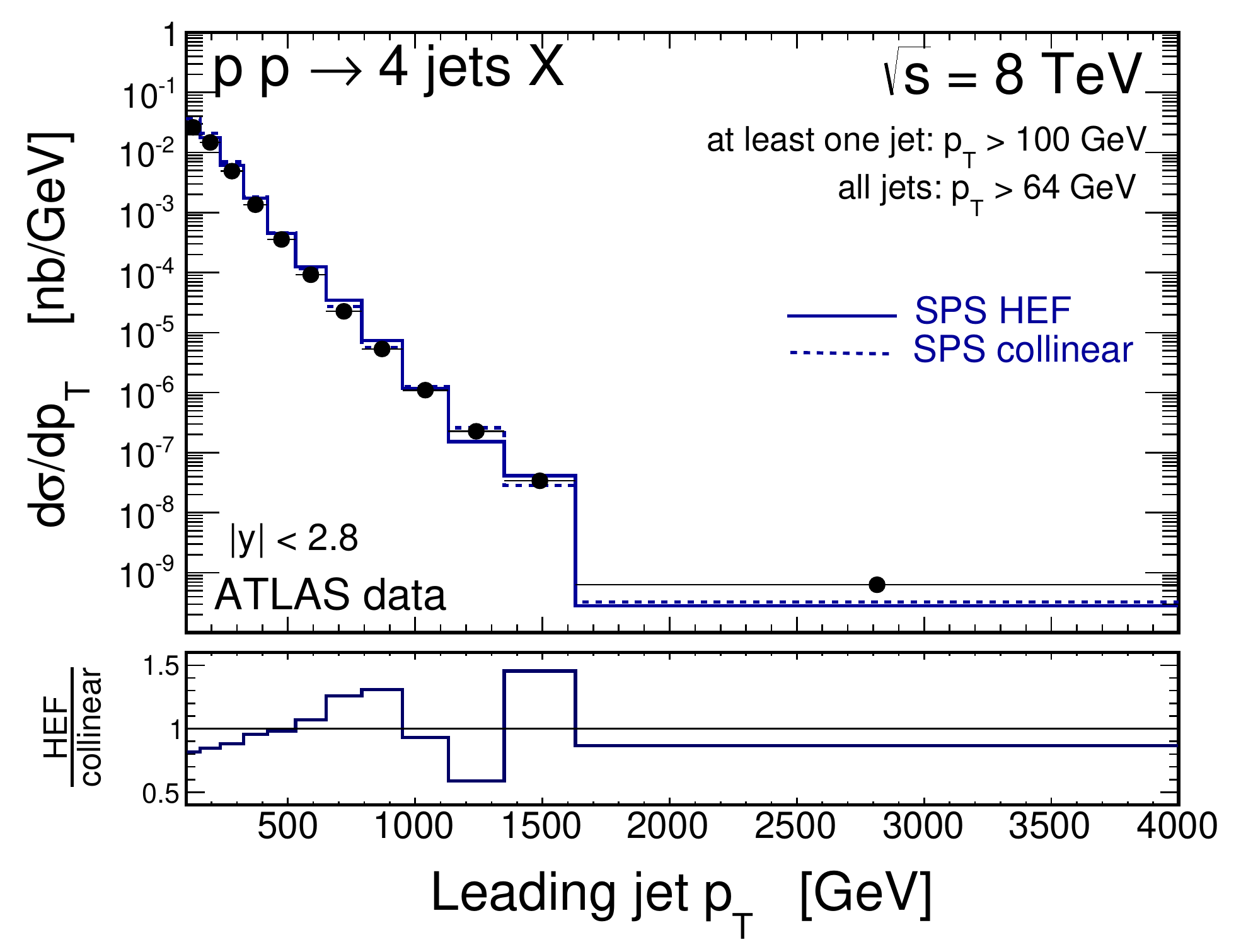}}
\end{minipage}
\hspace{0.5cm}
\begin{minipage}{0.47\textwidth}
 \centerline{\includegraphics[width=1.0\textwidth]{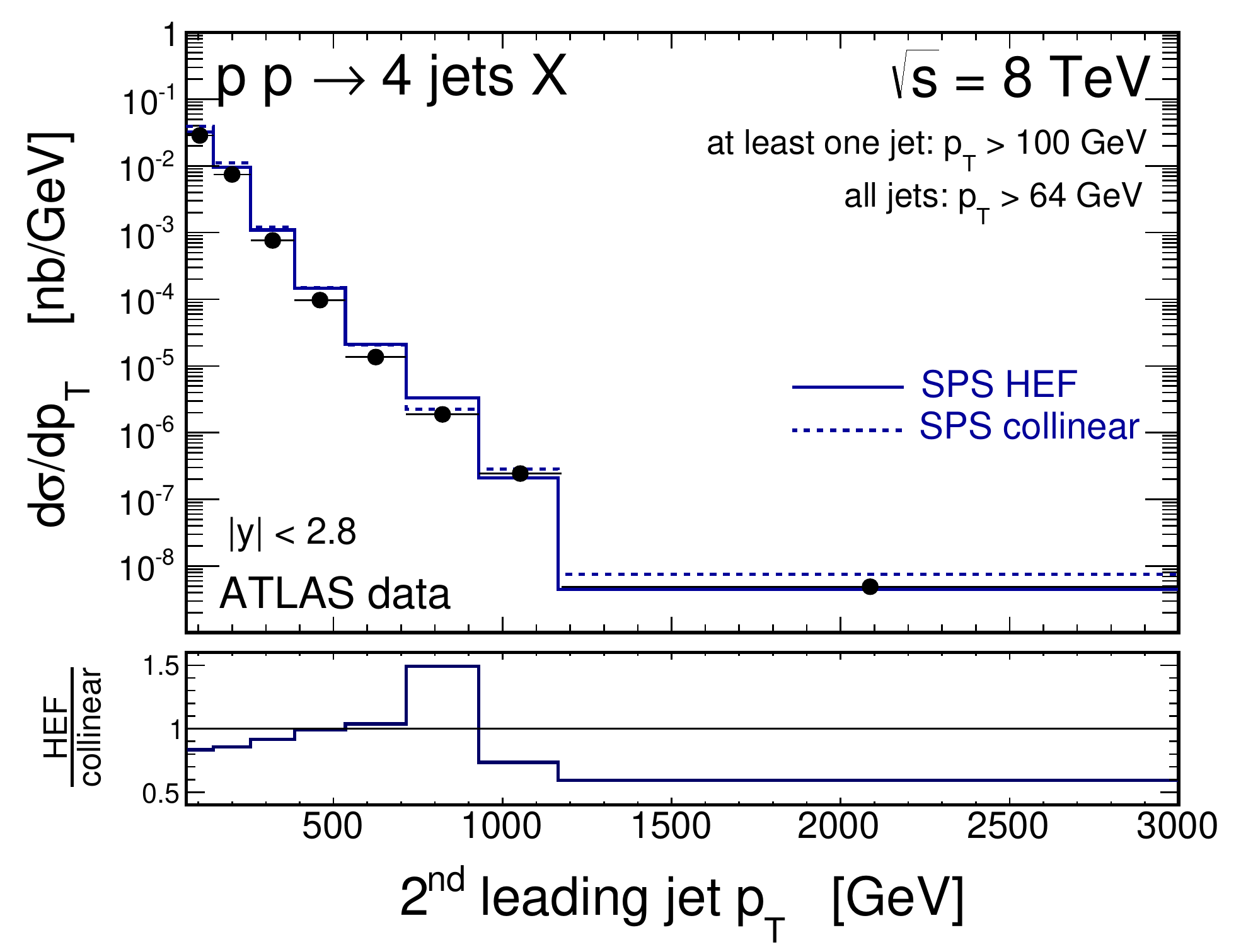}}
\end{minipage}
\caption{
Comparison of the HEF results to the collinear LO predictions and the ATLAS data for the 1st and 2nd leading jets.
In addition we show the ratio of the SPS HEF to the SPS collinear result.}
\label{Hard_central_ktvscoll1}
\end{figure}

\begin{figure}[h]
\begin{minipage}{0.47\textwidth}
 \centerline{\includegraphics[width=1.0\textwidth]{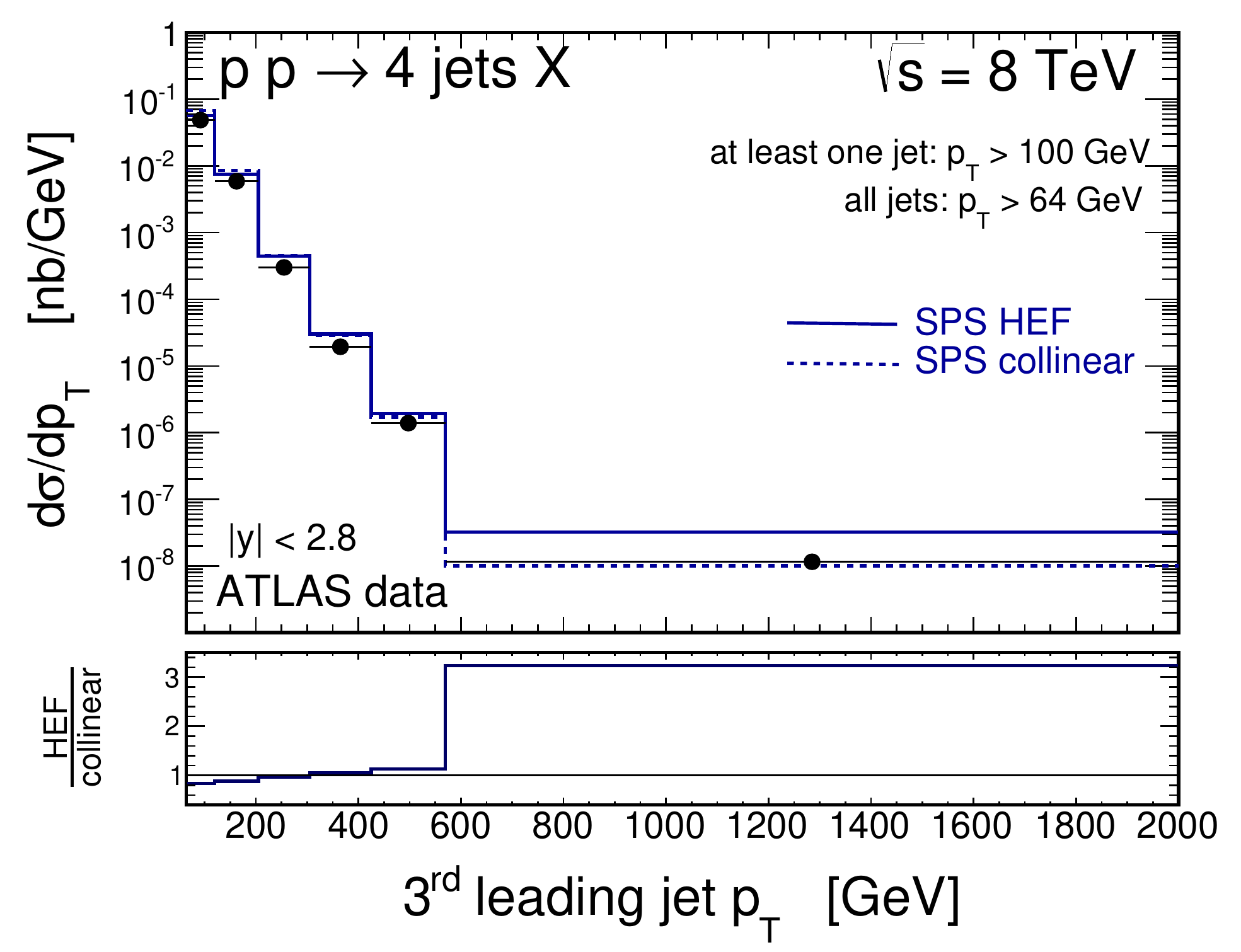}}
\end{minipage}
\hspace{0.5cm}
\begin{minipage}{0.47\textwidth}
 \centerline{\includegraphics[width=1.0\textwidth]{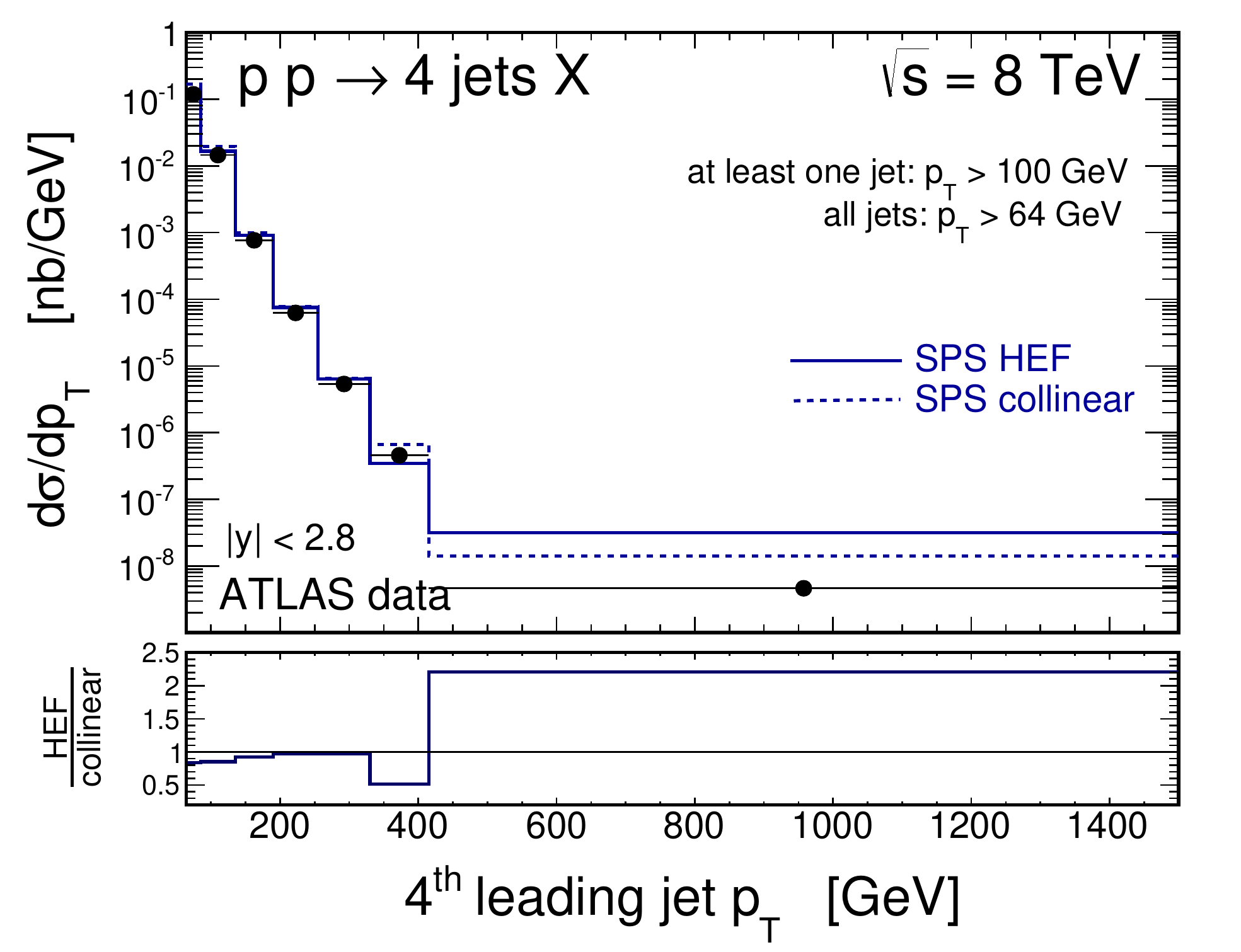}}
\end{minipage}
\caption{
Comparison of the HEF resuts to the collinear LO predictions and the ATLAS data for the 3rd and 4th leading jets.
In addition we show the ratio of the SPS HEF to the SPS collinear result.}
\label{Hard_central_ktvscoll2}
\end{figure}

Next we assess the difference between the HEF and collinear predictions at LO
as far as SPS is concerned.
We see from Figs.~\ref{Hard_central_ktvscoll1} and \ref{Hard_central_ktvscoll2} that
the collinear
factorization performs slightly better for intermediate values and HEF
does a better job for the last bins, except for the $4$th jet.
All in all, both approaches are consistent with the data in this kinematic region.
%
%

\subsection{Comparison to CMS data with softer cuts}

As discussed in Ref.~\cite{Maciula:2015vza}, so far the only experimental analysis 
of four-jet production relevant for the DPS studies was realized
by the CMS collaboration \cite{Chatrchyan:2013qza}. 
The cuts used in this analysis are $p_T> 50$ GeV for the first and
second jets, $p_T > 20$ GeV for the third and fourth jets,
$|\eta| < 4.7$ and the jet cone radius parameter $\Delta R > 0.5$. 
In the rest of this section, we present our results for such cuts.

As for the total cross section for the four jet production, the experimental and theoretical LO results are respectively:
\bea
\text{CMS collaboration} : &&
\sigma_{tot} = 330 \pm 5\, (\text{stat.}) \pm 45\, (\text{syst.})\,  nb
\nn \\
\text{LO collinear factorization}: &&
\sigma_{SPS} = 697\,  nb\, , \quad \sigma_{DPS} = 125\, nb \, ,  \quad \sigma_{tot} = 822\, nb 
\nn \\
\text{LO HEF $k_{T}$-factorization}: &&
\sigma_{SPS} = 548\,  nb\, , \quad \sigma_{DPS} = 33\, nb \, , \quad \sigma_{tot} = 581\, nb
\label{sigma_CMS}
\eea
It goes without saying that the LO result needs refinements from NLO contributions,
much more than it does in the case of the ATLAS hard cuts, 
as we are of course less deep into the perturbative region.
For this reason, in the following we will always perform comparisons only to 
data (re)normalised to the total (SPS+DPS) cross sections.
What is interesting in the HEF result, 
compared to collinear factorization, is the dramatic damping of the DPS
contribution. The effect of the damping is of kinematical nature and will be
explained below.

\begin{figure}[h]
\begin{center}
\begin{minipage}{0.47\textwidth}
 \centerline{\includegraphics[width=1.0\textwidth]{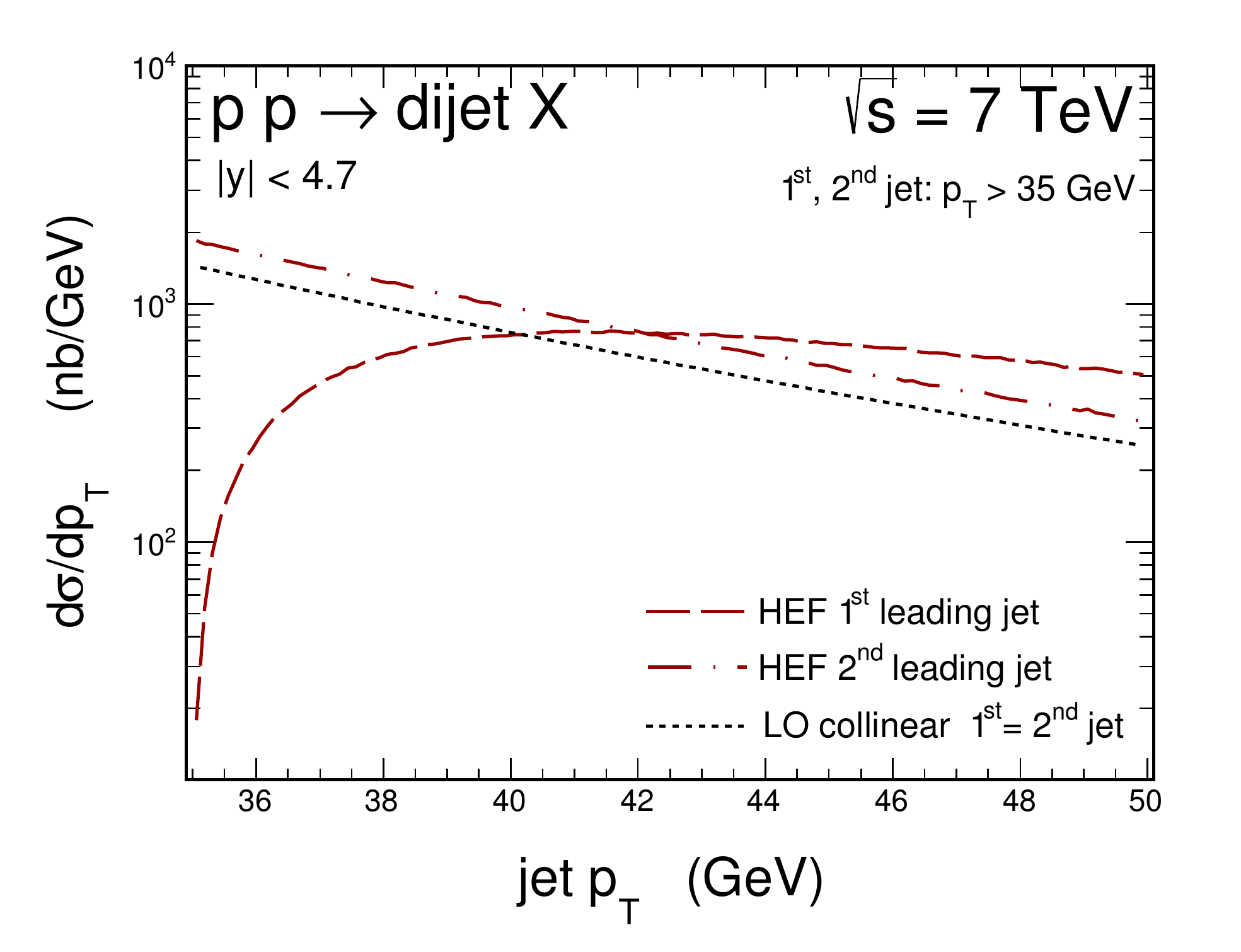}}
\end{minipage}
\end{center}
\caption{
The transverse momentum distribution of the leading (long dashed line) 
and subleading (long dashed-dotted line) jet for the dijet production in HEF. 
For comparison we show also result for leading-order collinear
approach (short dashed line) in which case both jets give 
the same distribution. 
}
\label{fig:illustration_of_damping}
\end{figure}

The effect of the relative damping of the HEF DPS result
compared to leading-order collinear DPS result is of kinematical origin.
The main idea can be understood already in a bit simpler case of two-jet
production within the HEF approach.
For the purpose of this illustration we impose a cut $p_{T} >$ 35 GeV on both jets (leading and
subleading).
In Fig.~~\ref{fig:illustration_of_damping} we show transverse momentum 
distribution for both leading (long-dashed line) and subleading (long dashed-dotted line)
jet. We observe a minimum for the leading jet and maximum for the subleading
jet for transverse momenta in the vicintity of the lower cut. 
The integrated cross section for the leading and subleading jet
is of course identical as they are "measured" (identified) in coincidence.
For the leading order collinear case both jets have the same distribution
and one gets maximum in the vicinity of the transverse momentum
threshold in both cases. In this case imposing cuts on both jets does not
lead to "loosing" cross section. 
In contrast, in the HEF approach, if the leading jet is close to the transverse momentum threshold, 
then the subleading jet is typically below that threshold, therefore such an event is not counted.

For four-jet DPS production the situation is more complicated and
strongly depends on cuts (all identical, two pairs of identical cuts, harder
cut for the leading jet and identical for the other, etc.).

%
\begin{figure}[h]
\begin{center}
\begin{minipage}{0.47\textwidth}
 \centerline{\includegraphics[width=1.0\textwidth]{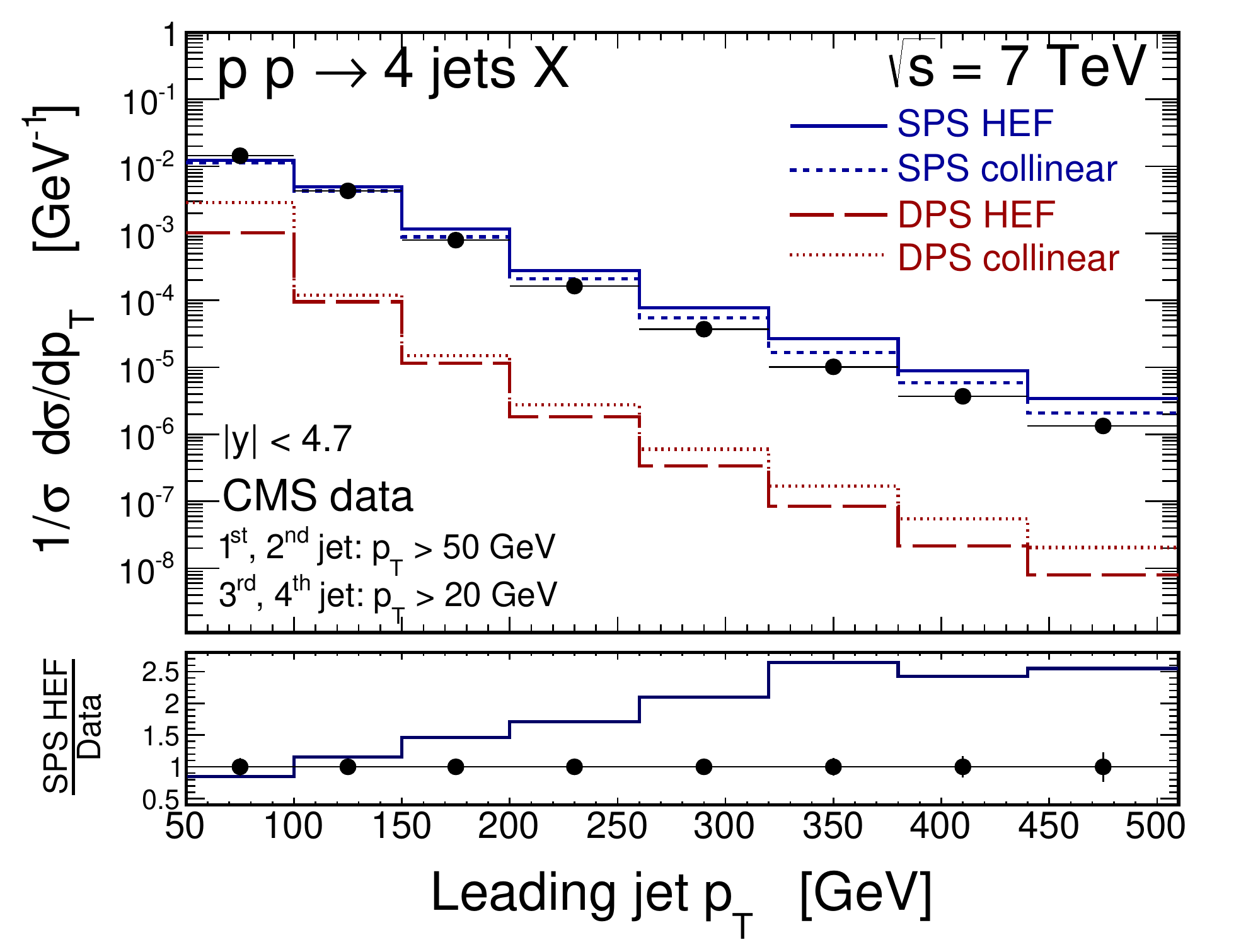}}
\end{minipage}
\hspace{0.5cm}
\begin{minipage}{0.47\textwidth}
 \centerline{\includegraphics[width=1.0\textwidth]{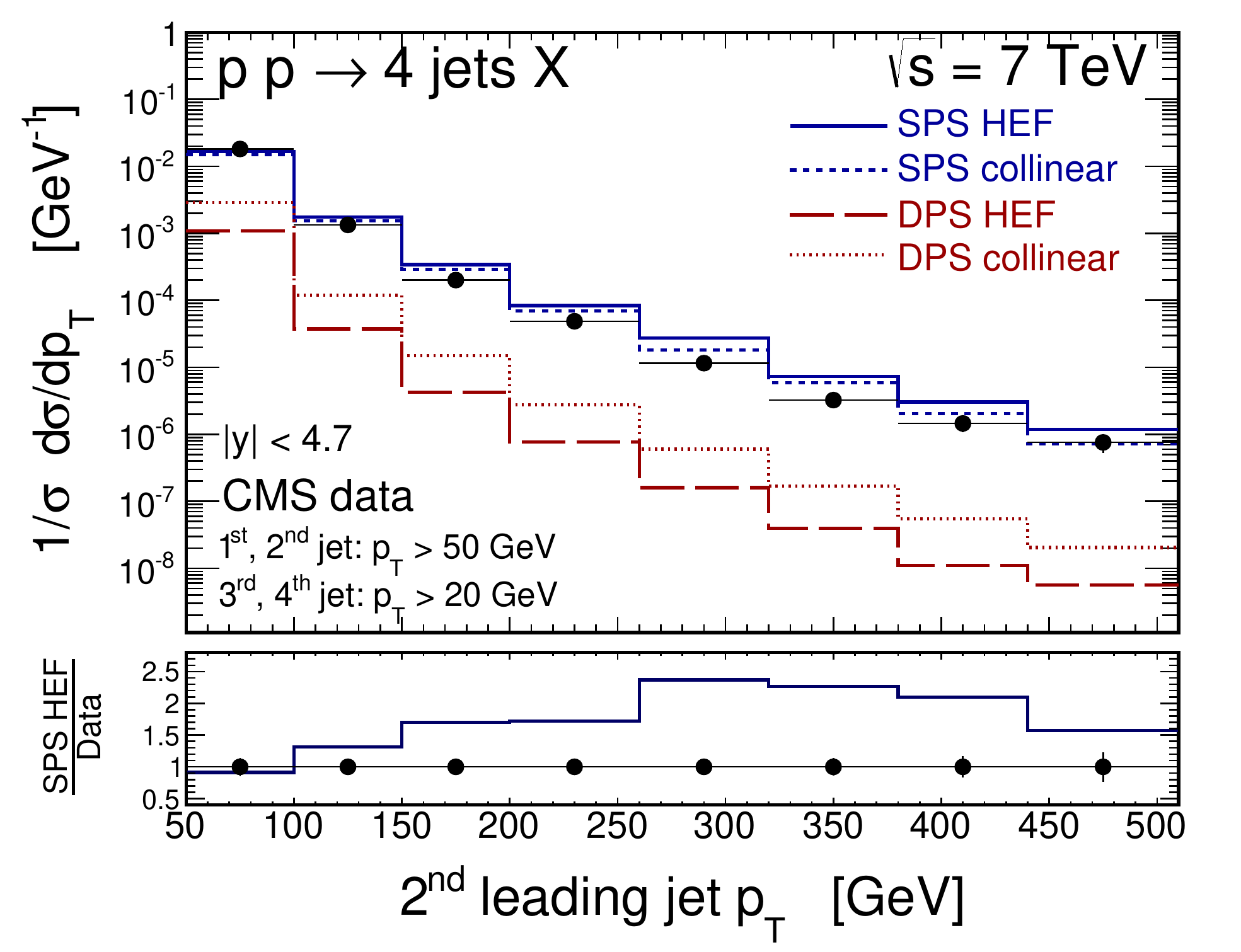}}
\end{minipage}
\end{center}
\caption{
Comparison of the LO collinear and HEF predictions to
the CMS data for the 1st and 2nd leading jets. In addition we show the ratio of the SPS HEF result to the CMS data.}
\label{CMS_pT_12}
\end{figure}
%
\begin{figure}[h]
\begin{center}
\begin{minipage}{0.47\textwidth}
 \centerline{\includegraphics[width=1.0\textwidth]{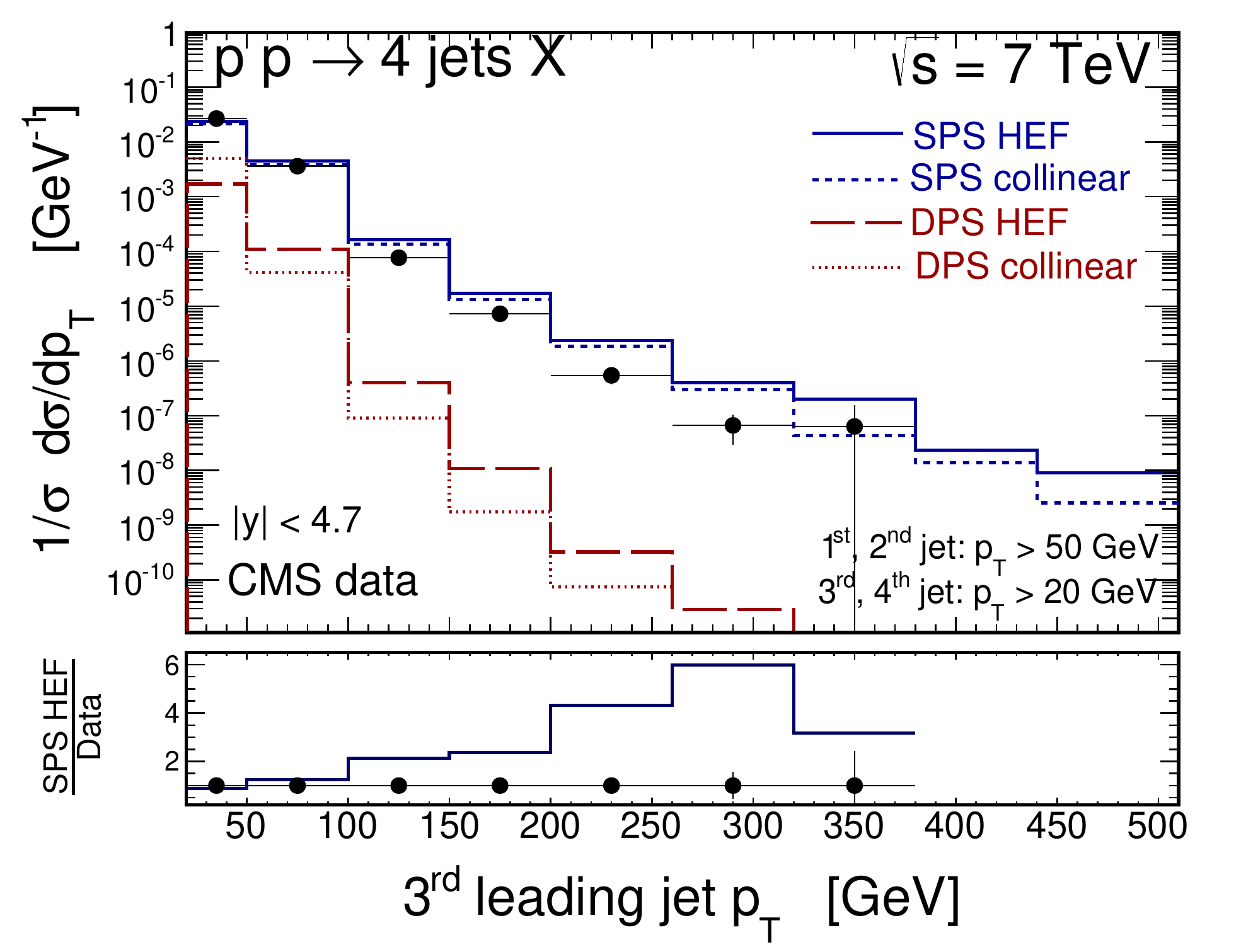}}
\end{minipage}
\hspace{0.5cm}
\begin{minipage}{0.47\textwidth}
 \centerline{\includegraphics[width=1.0\textwidth]{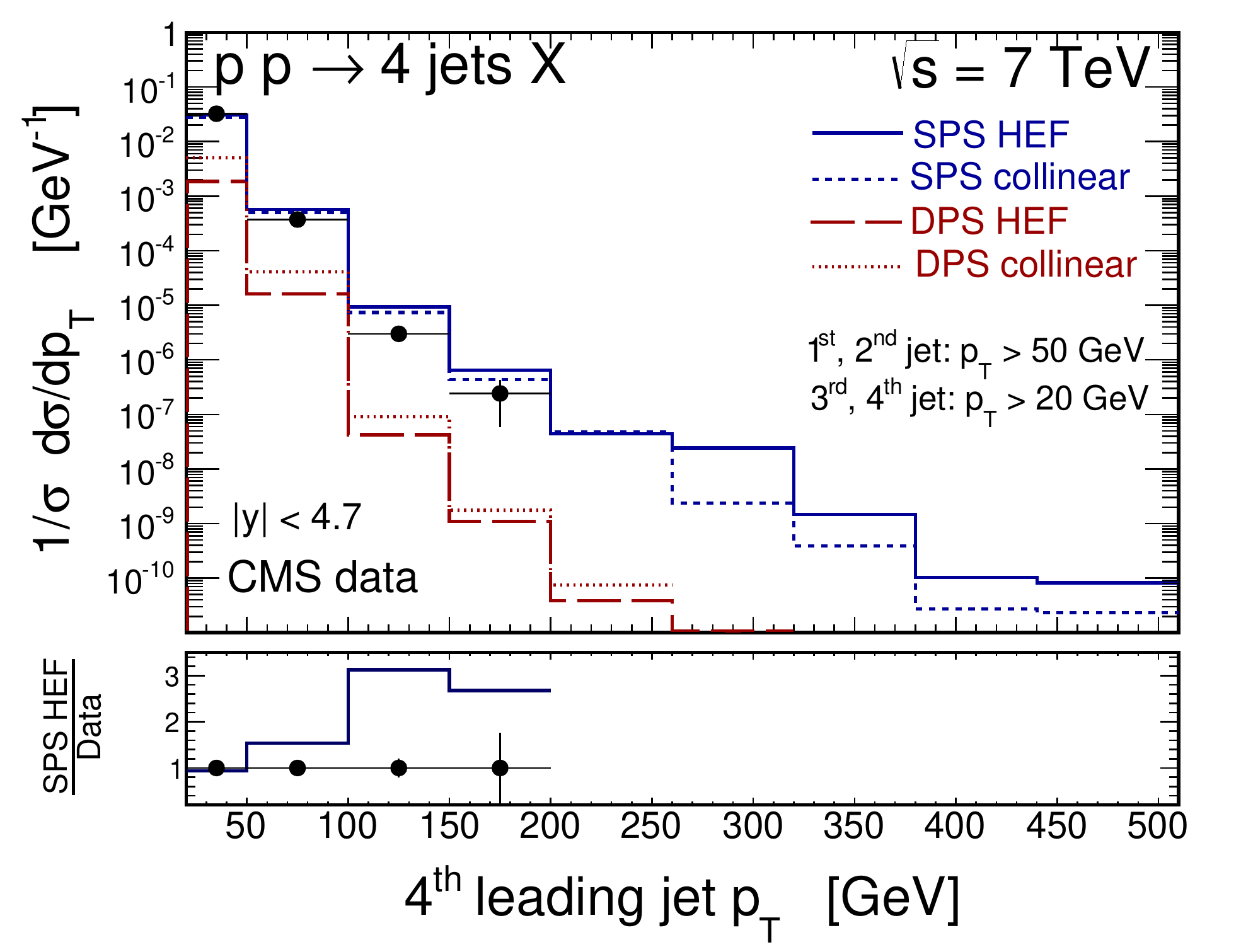}}
\end{minipage}
\end{center}
\caption{
Comparison of the LO collinear and HEF predictions to 
the CMS data for the 3rd and 4th leading jets. In addition we show the ratio of the SPS HEF result to the CMS data.}
\label{CMS_pT_34}
\end{figure}
%
In Figs.~\ref{CMS_pT_12} and \ref{CMS_pT_34} we compare the predictions
in HEF to the CMS data. Here both the SPS and DPS
contributions are normalized to the total cross section,
i.e. the sum of the SPS and DPS contributions.
In all cases the renormalized transverse momentum distributions agree
with the CMS data. However, the absolute cross sections obtained in this case
within the HEF approach are too large.

Not only transverse momentum dependence is interesting.
The CMS collaboration extracted also a more complicated observables
\cite{Chatrchyan:2013qza}.
One of them, which involves all four jets in the final state,
is the $\Delta S$ variable, defined in Ref.~\cite{Chatrchyan:2013qza} as the angle between pairs of
the harder and the softer jets,
\beq
\Delta S = \arccos \left( 
\frac{\vec{p}_T(j^{\text{hard}}_1,j^{\text{hard}}_2) \cdot \vec{p}_T(j^{\text{soft}}_1,j^{\text{soft}}_2)}{|\vec{p}_T(j^{\text{hard}}_1,j^{\text{hard}}_2)|\cdot|\vec{p}_T(j^{\text{soft}}_1,j^{\text{soft}}_2)|}  
\right) \, ,
\eeq
where $\vec{p}_T(j_i,j_k)$ stands for the sum of the transverse momenta of the two jets in arguments.

\begin{figure}[h]
\begin{center}
\begin{minipage}{0.47\textwidth}
 \centerline{\includegraphics[width=1.0\textwidth]{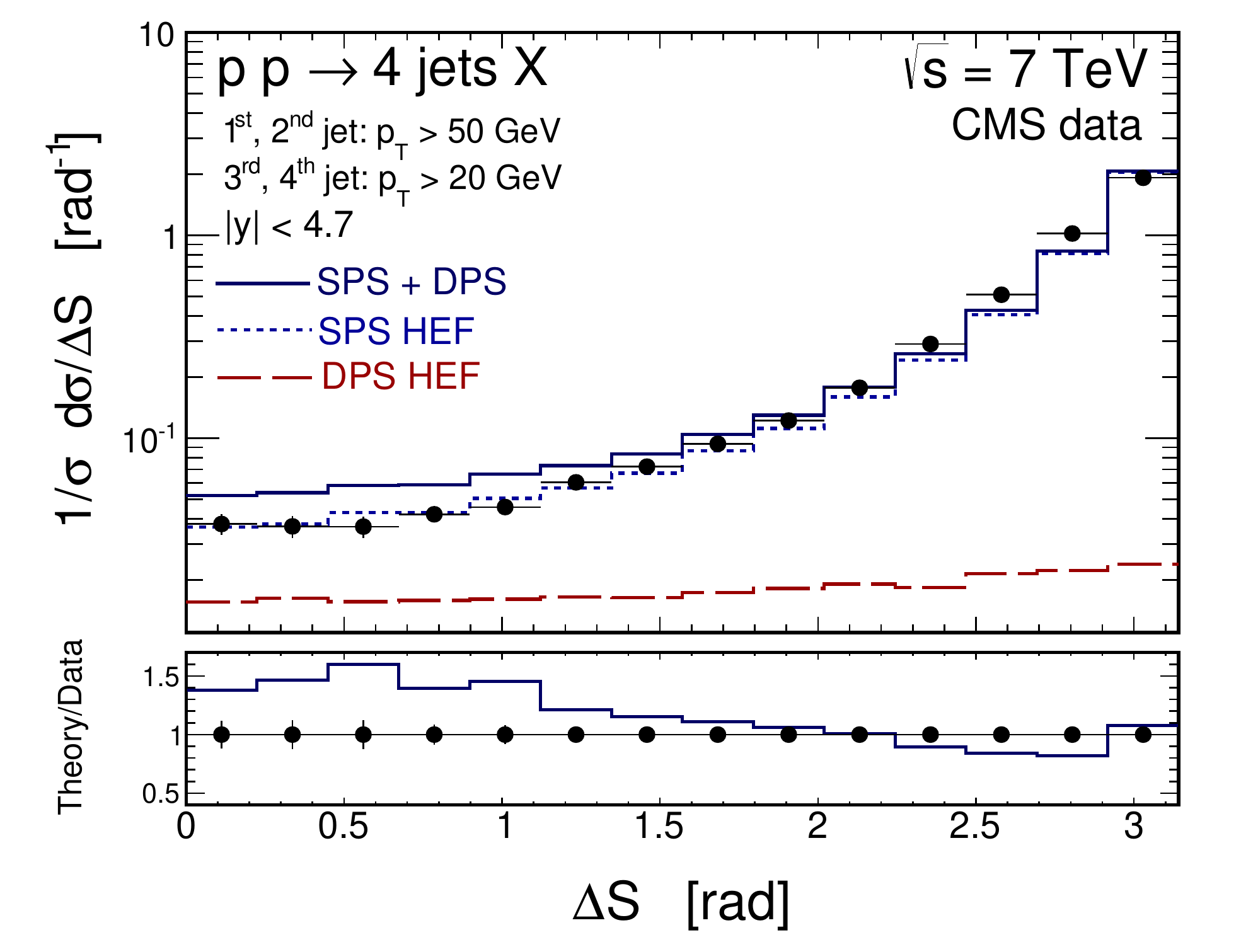}}
\end{minipage}
\end{center}
\caption{
Comparison of the HEF predictions to the CMS data for $\Delta S$ spectrum. In addition we show the ratio of the (SPS+DPS) HEF result to the CMS data.}
\label{fig:CMS_DS}
\end{figure}

In Fig.~ \ref{fig:CMS_DS} we present our HEF prediction for
the normalized to unity distribution in the $\Delta S$ variable. 
Our HEF result approximately agrees with the
experimental $\Delta S$ distribution. 
In contrast the LO collinear approach leads to 
$\Delta S$ = 0, i.e. a Kronecker-delta peak at $\Delta S$ = 0 for 
the distribution in $\Delta S$.
For the DPS case this is rather trivial. The two hard and two soft 
jets come in this case from the same scatterings and are back-to-back
(LO), so each term in the argument of $arccos$ is zero 
(jets are balanced in transverse momenta).
For the SPS case the transverse momenta of the two jet pairs 
(with hard jets and soft jets) are identical and have opposite direction
(the total transverse momentum of all four jets must be zero from the momentum conservation).
Then it is easy to see that the argument of $arccos$ is just -1.
This means $\Delta S$ = 0.
The above relations are not fullfilled in the HEF
approach. The SPS contribution clearly dominates and approximately gives the shape
of the $\Delta S$ distribution.
The DPS contribution improves the agreement with the data in the 
central region, worsening it a little bit for $\Delta S \rightarrow 0$
while essentially leaving the result unaffected for $\Delta S
\rightarrow \pi$.
It is anyway interesting that we roughly describe the data
via pQCD effects within our HEF approach which 
are in Ref.~\cite{Chatrchyan:2013qza} described by parton-showers and soft MPIs. 

%
%

\begin{figure}[h]
\begin{center}
\begin{minipage}{0.47\textwidth}
 \centerline{\includegraphics[width=1.0\textwidth]{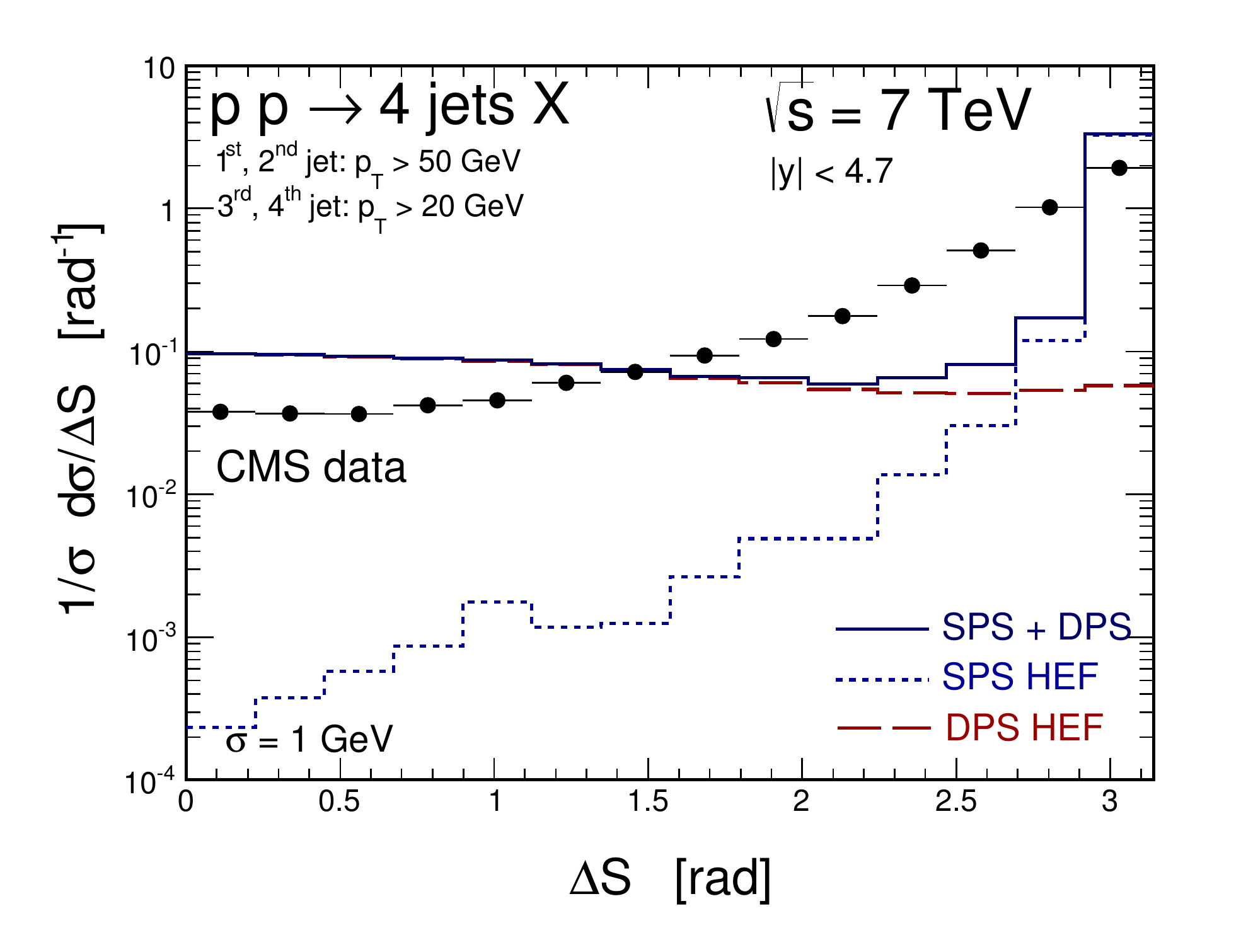}}
\end{minipage}
\hspace{0.5cm}
\begin{minipage}{0.47\textwidth}
 \centerline{\includegraphics[width=1.0\textwidth]{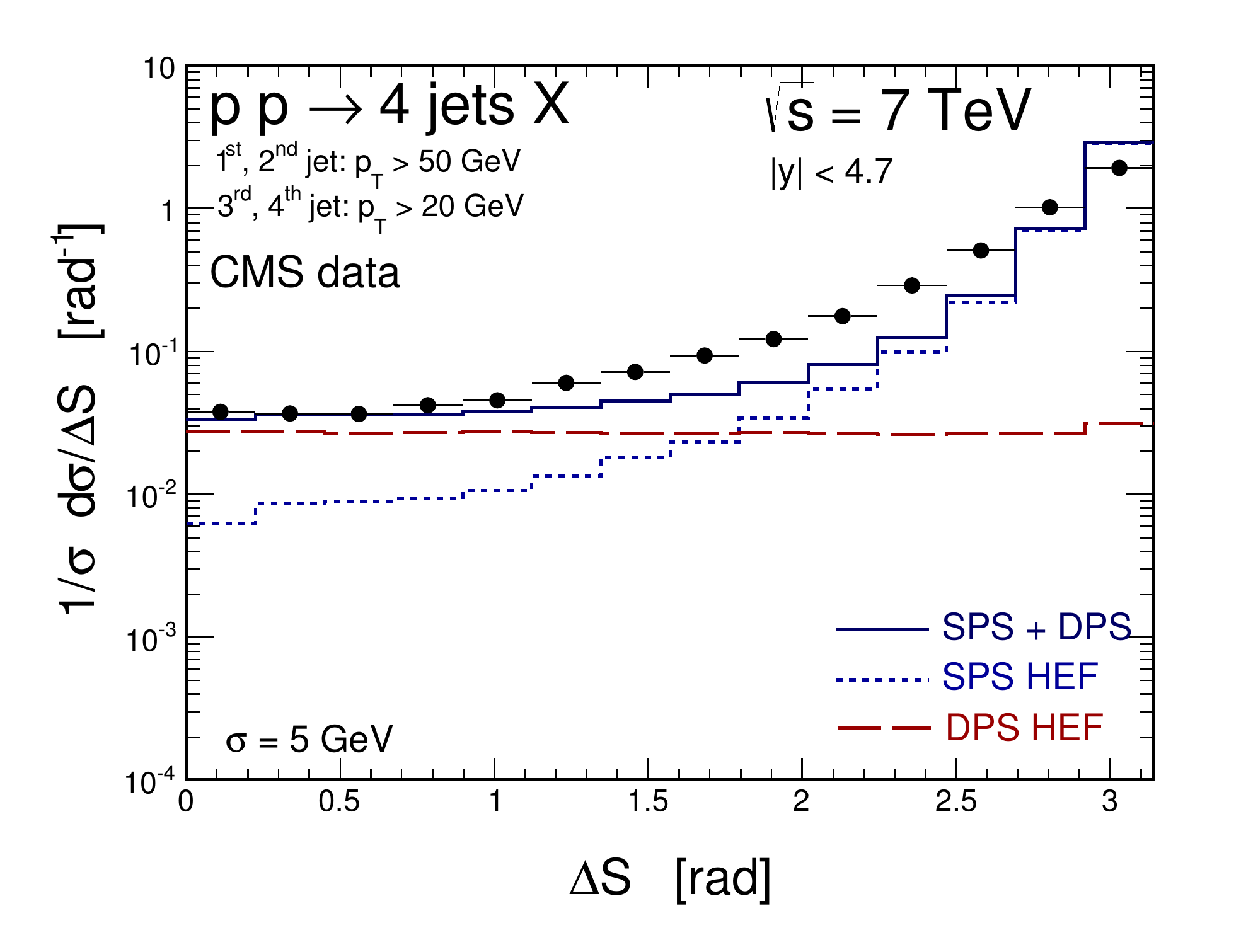}}
\end{minipage}
\end{center}
\caption{
Distribution in $\Delta S$ for the toy Gaussian model of TMDs
with $\sigma$ = 1 GeV (left) and $\sigma$ = 5 GeV (right).
}
\label{fig:CMS_DS_toymodel}
\end{figure}

It would be nice to have more insight into our successful
description of the $\Delta S$ distribution measured by the CMS experiment.
In Fig.~\ref{fig:CMS_DS_toymodel} we return to the $\Delta S$ spectrum and show also
two results with a TMD toy model with the Gaussian smearing of 
the collinear parton distribution:
\begin{equation}
{\cal F}_p(x,k_T^2,\mu^2) = G(k_T^2;\sigma) x p (x,\mu^2).
\label{Gaussian_UGDF}
\end{equation}
We take two sets of smearing parameter: $\sigma$ = 1 GeV (left panel) and 5 GeV (right panel).
Taking a bigger value of $\sigma$ we approach the CMS data. This shows that the transverse momenta bigger than a few GeV are
needed to approach the data. The disagreement of the toy model with $\sigma = 5$ GeV result with the experimental data and the agreement for the 
DLC-2016 model illustrate sensitivity to TMD's. 

%
%

\subsection{HEF predictions for a possible set of asymmetric cuts }

Moving from the previous considerations, in the following subsection we
present our results of the DPS employing asymmetric cuts by which we
mean here $p_T > 35$ GeV for the leading jet, 
$p_T > 20$ GeV for the other jets and $|\eta| < 4.7$, $\Delta R > 0.5$.
Of course it would be interesting to have the results of such an experimental analysis ( i.e. with soft enough
but asymmetric cuts ) in order to test the predictions of HEF for DPS.

In this case the theoretical total cross sections for four-jet production are:
\bea
\text{LO collinear factorization}: &&
\sigma_{SPS} = 1969 \,  nb\, , \quad \sigma_{DPS} = 514\, nb \, ,  \quad \sigma_{tot} = 2309\, nb 
\nn \\
\text{LO HEF $k_{T}$-factorization}: &&
\sigma_{SPS} = 1506 \,  nb\, , \quad \sigma_{DPS} =  297 \, nb \, , \quad \sigma_{tot} = 1803\, nb
\eea
Compared to (\ref{sigma_CMS}), it is apparent that now the drop in the total cross section for DPS 
when moving from LO collinear to HEF approach is considerably
smaller. Here we have enough phase space for the subleading jet(s),
as a consequence of the asymmetric cuts.

%
\begin{figure}[h]
\begin{center}
\begin{minipage}{0.47\textwidth}
 \centerline{\includegraphics[width=1.0\textwidth]{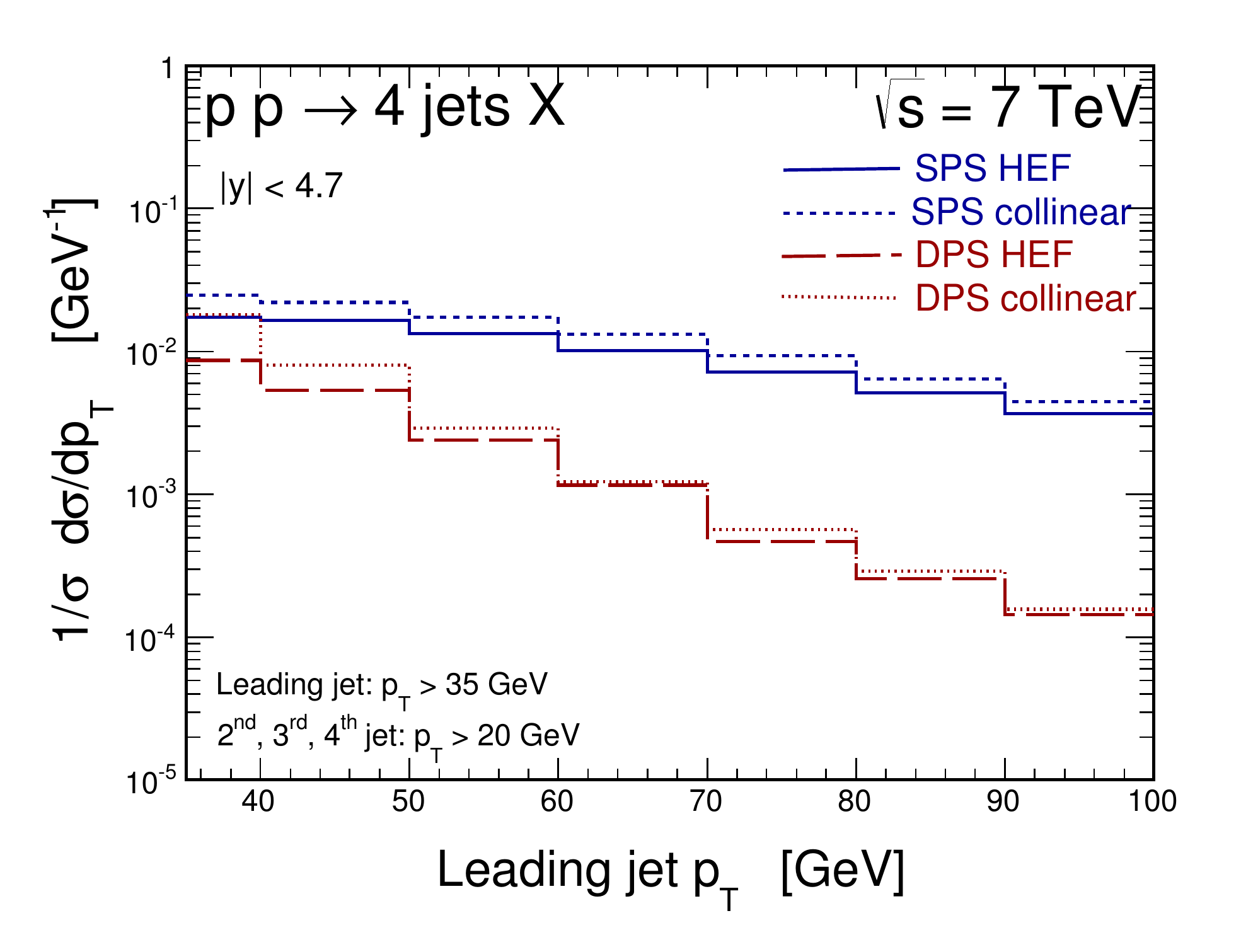}}
\end{minipage}
\hspace{0.5cm}
\begin{minipage}{0.47\textwidth}
 \centerline{\includegraphics[width=1.0\textwidth]{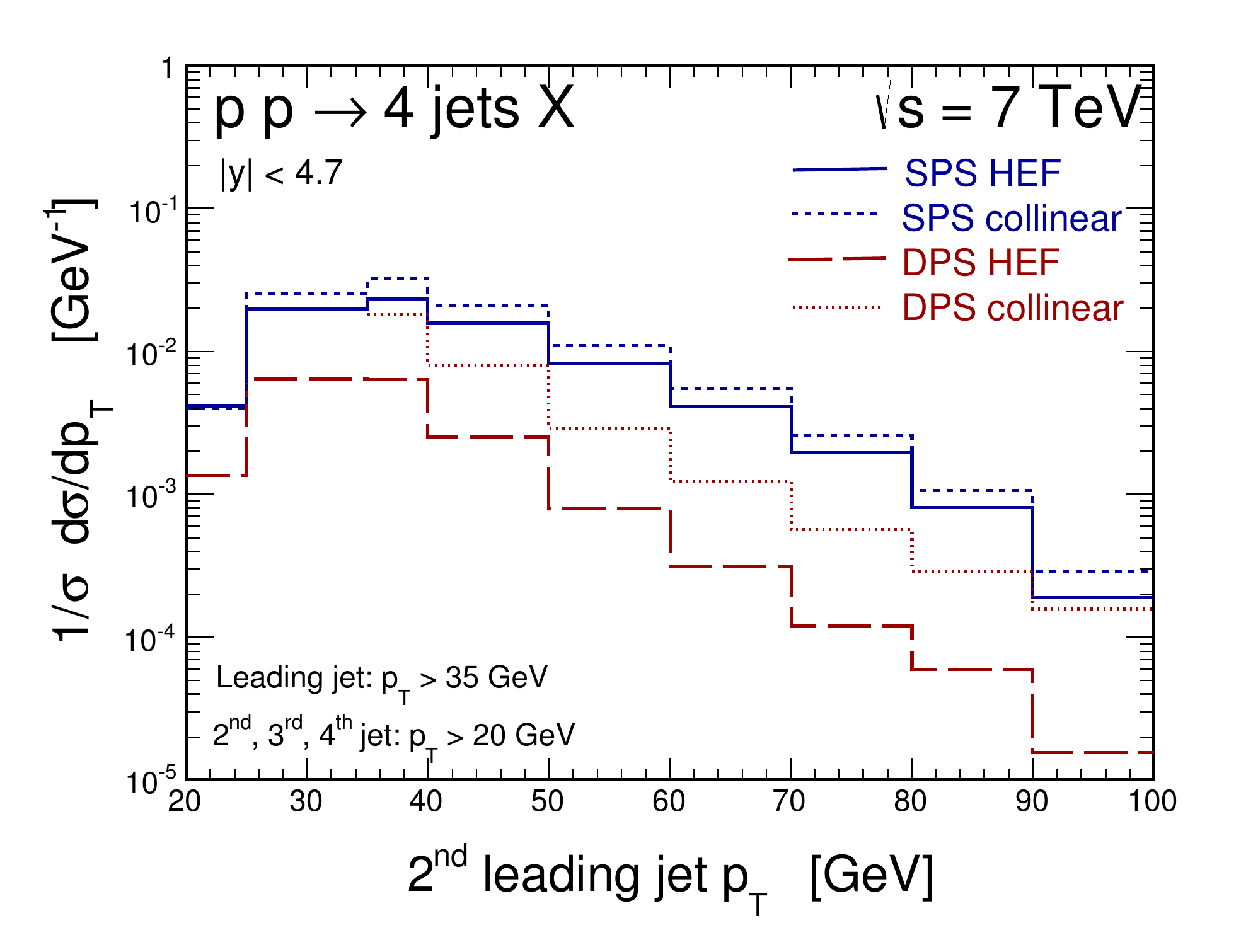}}
\end{minipage}
\end{center}
\caption{
LO collinear and HEF predictions for the 1st and 2nd 
leading jets with the asymmetric cuts.}
\label{Asymm_pT_12}
\end{figure}
%
\begin{figure}[h]
\begin{center}
\begin{minipage}{0.47\textwidth}
 \centerline{\includegraphics[width=1.0\textwidth]{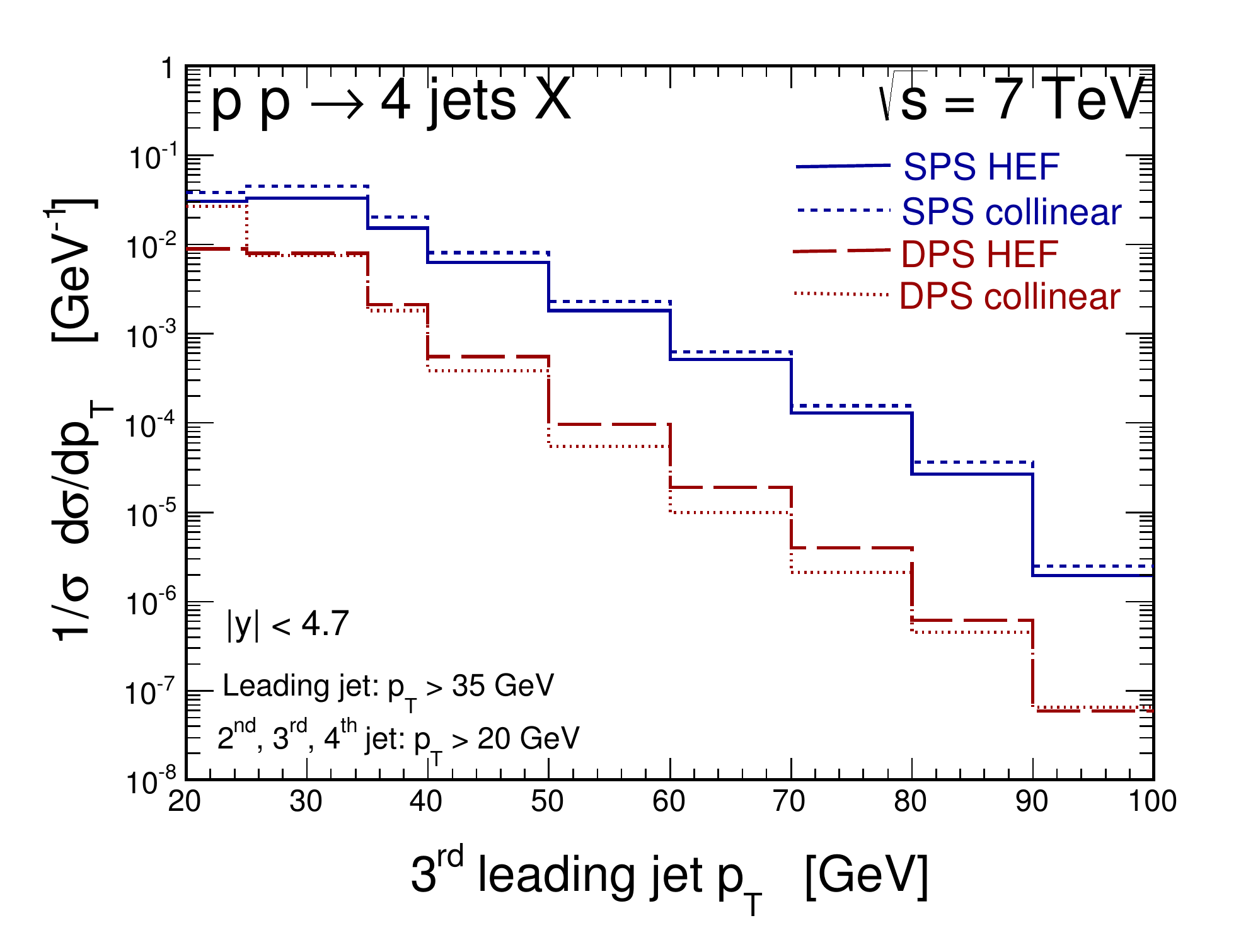}}
\end{minipage}
\hspace{0.5cm}
\begin{minipage}{0.47\textwidth}
 \centerline{\includegraphics[width=1.0\textwidth]{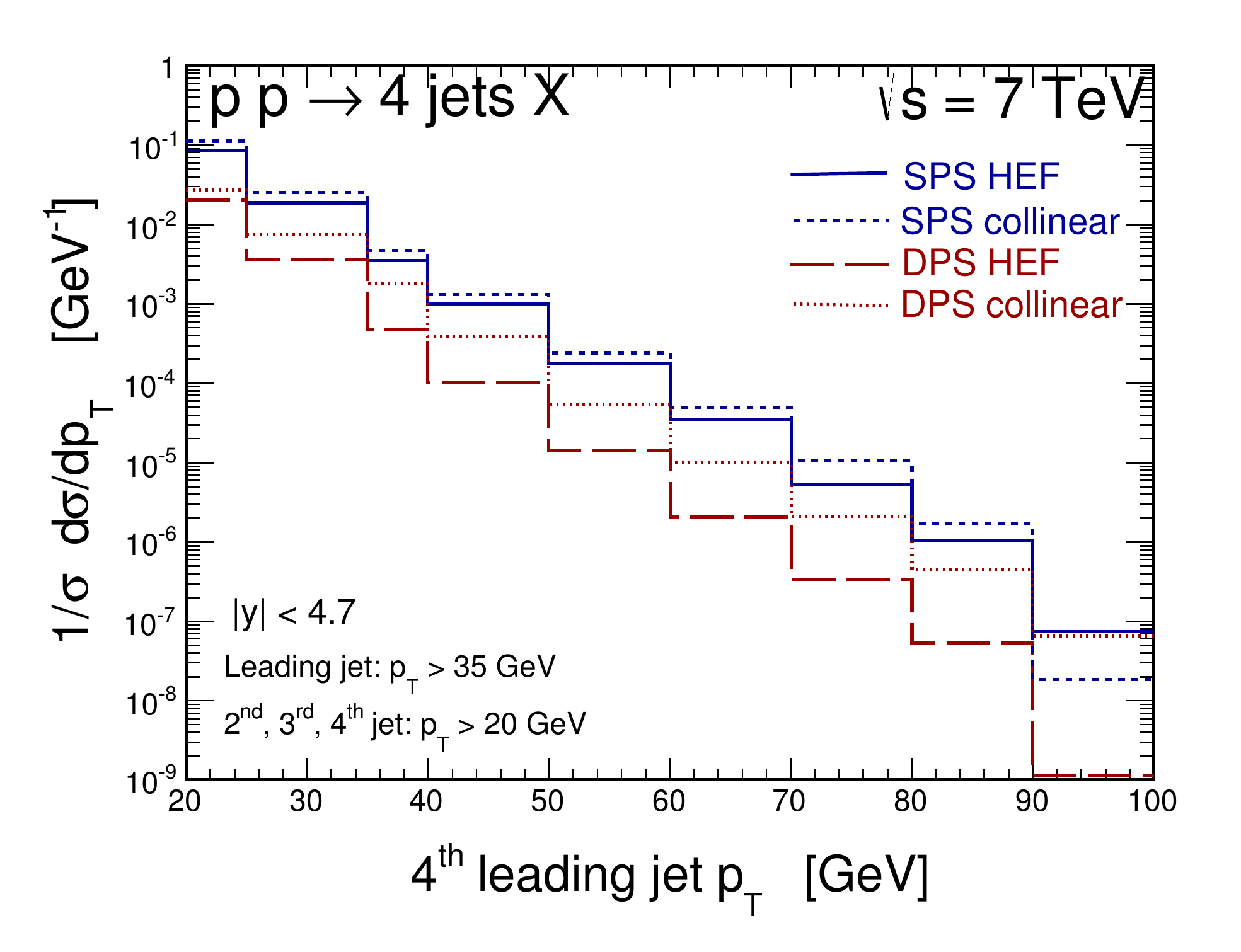}}
\end{minipage}
\end{center}
\caption{
LO collinear and HEF predictions for the 3rd and 4th 
leading jets with asymmetric cuts.}
\label{Asymm_pT_34}
\end{figure}
%

In Figs.~\ref{Asymm_pT_12} and \ref{Asymm_pT_34} we show our predictions
for the normalized transverse momentum distributions with the new set of cuts.

\begin{figure}[h]
\begin{center}
\begin{minipage}{0.47\textwidth}
 \centerline{\includegraphics[width=1.0\textwidth]{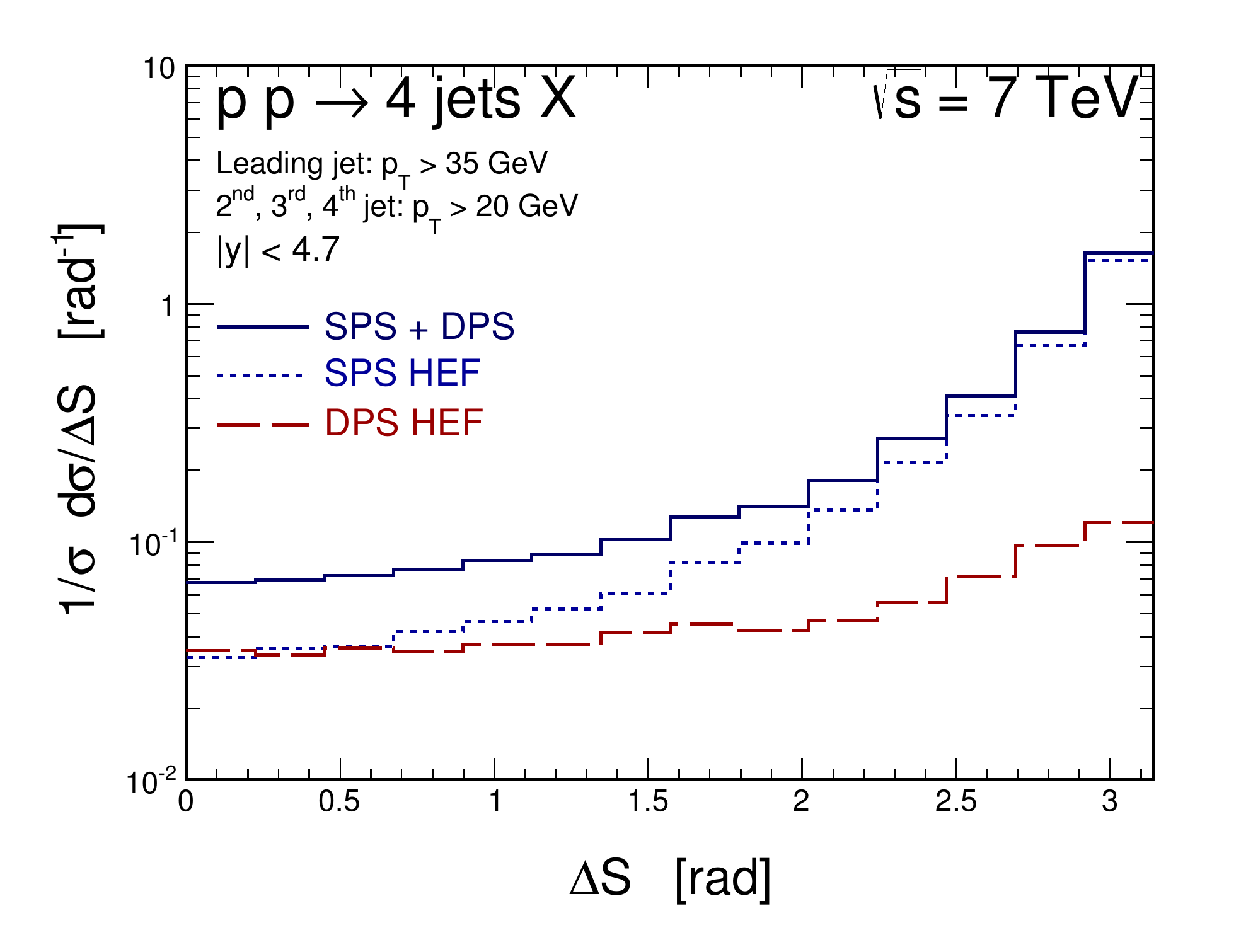}}
\end{minipage}
\end{center}
\caption{
HEF prediction for $\Delta S$ with asymmetric cuts.}
\label{Asymm_DS}
\end{figure}
%

\section{Conclusions}

In the present paper we have compared the perturbative predictions 
for four-jet production at the LHC in leading-order collinear and
high-energy ($k_T$-)factorization.
Both single-parton scattering and double parton contribution 
have been calculated for a first time in the high-energy
($k_T$-)factorization approach. 
The calculation of the SPS contribution may be considered as a technical
achievment. So far only production of the $c \bar c c \bar c$ final
state (also of the 2 $\to$ 4 type) was discussed in the literature in this context.
For the four-jet production the number of relevant subproceses is 
much larger but could be treated in our automatized framework.

We find that both collinear and the ($k_T$-)factorization approaches
describe the data for hard central cuts, relevant for 
the ATLAS experiment, reasonably well. For the harder cuts we get both
normalization and shape of the transverse momentum distributions. For the softer cuts
used e.g. by the CMS collaboration the tree level result is unreliable.
Therefore in this case we have presented results for normalised cross
sections.
We have presented distributions in transverse momenta for all jets
ordered in their transverse momenta.
We have found that for symmetric cuts the DPS cross section obtained
with more realistic high energy ($k_T$-)factorization approach is smaller than the one
obtained in the collinear approach. This is important result in searches
for DPS effects in four-jet production not discussed so far in the
literature. We have tried to explain this as kinematical effect due to
phase space limitation when simultaneously imposing cuts on all jets 
but a full explanation is a bit intricate.
While we observe, in agreement with Ref.~\cite{Maciula:2015vza}, that lowering the cut 
in transverse momenta can significantly enhance the experimental sensitivity to DPS, 
we also observe that the HEF approach severely tames this effect for symmetric cuts,
due to gluon-emission effects which alter the transverse-momentum balance between final state partons. 
We have found that the damping is not present when cuts are not identical.
The discussion how to optimize the cuts will be presented elsewhere.

For other approaches addressing the four-jet production and resummation 
of BFKL type of singularities \cite{Vera:2007kn,Vera:2007dr} we refer 
the Reader to Ref.~\cite{Caporale:2015int}. The authors of Ref.~\cite{Caporale:2015int}
define new angular four jet observables to test BFKL approach.

As a side result, we present in Appendix \ref{Comparison} 
a detailed numerical comparison of results obtained for dijet production with matrix element generated
automatically by means of AVHLIB with those obtained analytically within the Parton-Reggeization-Approach (PRA) in
Ref.~\cite{Nefedov:2013ywa} and implemented in an independent code. 
For all (sub)processes we have obtained very good agreement of corresponding differential distributions. 
We show corresponding azimuthal angle correlations for different subprocesses which are particularly 
efficient for such tests, as they sample the situation in a broad range of the phase space.
It was shown in the past for some subprocesses that also analytical results coincide \cite{vanHameren:2013csa}.

\section*{Acknowledgments}
The work of M.S. and K.K. have been supported by Narodowe Centrum Nauki
with Sonata Bis grant DEC-2013/10/E/ST2/00656 while R.M. and A.S.
have been supported by the Polish National Science Center grant
DEC-2014/15/B/ST2/02528. A.v.H. was supported by
grant of National Science Center, Poland, No. 2015/17/B/ST2/01838.
M.S. also thanks the "Angelo della Riccia" foundation for support.
We wish to thank A. V. Shipilova for useful discussion during the development of this project
on analytic formula for matrix elements. 

\appendix
\section{Construction of the TMDs}\label{App_TMDs}

%
The Kimber-Martin-Ryskin (KMR) analysis of coherence effects provides in the
DLL limit the following formula for obtaining TMD parton density
function (this limit is relevant for us since we consider moderate
values of $x$ and rather large scales):
\be
{\cal F}_i(x,k_{T}^2,\mu^2)=\theta(k_{T}^2-k_{T\,min}^2)\theta(\mu^2-\mu^{2}_{min})\partial_{k_{T}^2}\left[xf_i(x,\kappa^2)T_i(\kappa,\mu)\right]|_{\kappa=k_{T}} ,
\ee
where $k_{T\,min}^{2} = \mu^{2}_{min} = 1.7$ GeV$^{2}$ and $T_i$ is appropriate Sudakov form factor. In gluon channel it reads
\be
T_g(k_T,\mu)=\exp\left(-\int_{k_T^2}^{\mu^2}\frac{\alpha_s(\mu^2)}{2\pi}\int_0^{\frac{\mu}{\mu+p_t}}dz^\prime(z^\prime P_{gg}(z^\prime)+n_F P_{qg}(z^\prime))\right)
\ee
while for quarks we have:
\be
T_q(k_T,\mu)=\exp\left(-\int_{k_T^2}^{\mu^2}\frac{\alpha_s(\mu^2)}{2\pi}\int_0^{\frac{\mu}{\mu+p_t}}dz^\prime P_{qq}(z^\prime) \right).
\ee
In our calculations we take for $xf_i$ CTEQ10NLO set of pdfs, the
$P_{ij}$ are LO splitting functions which to accuracy we work with are
sufficient and in practical calculations we take $n_F$ = 5.

\section{Comparison of $2 \to 2$ matrix elements}\label{Comparison}

\begin{figure}[!h]
\begin{minipage}{0.47\textwidth}
 \centerline{\includegraphics[width=1.0\textwidth]{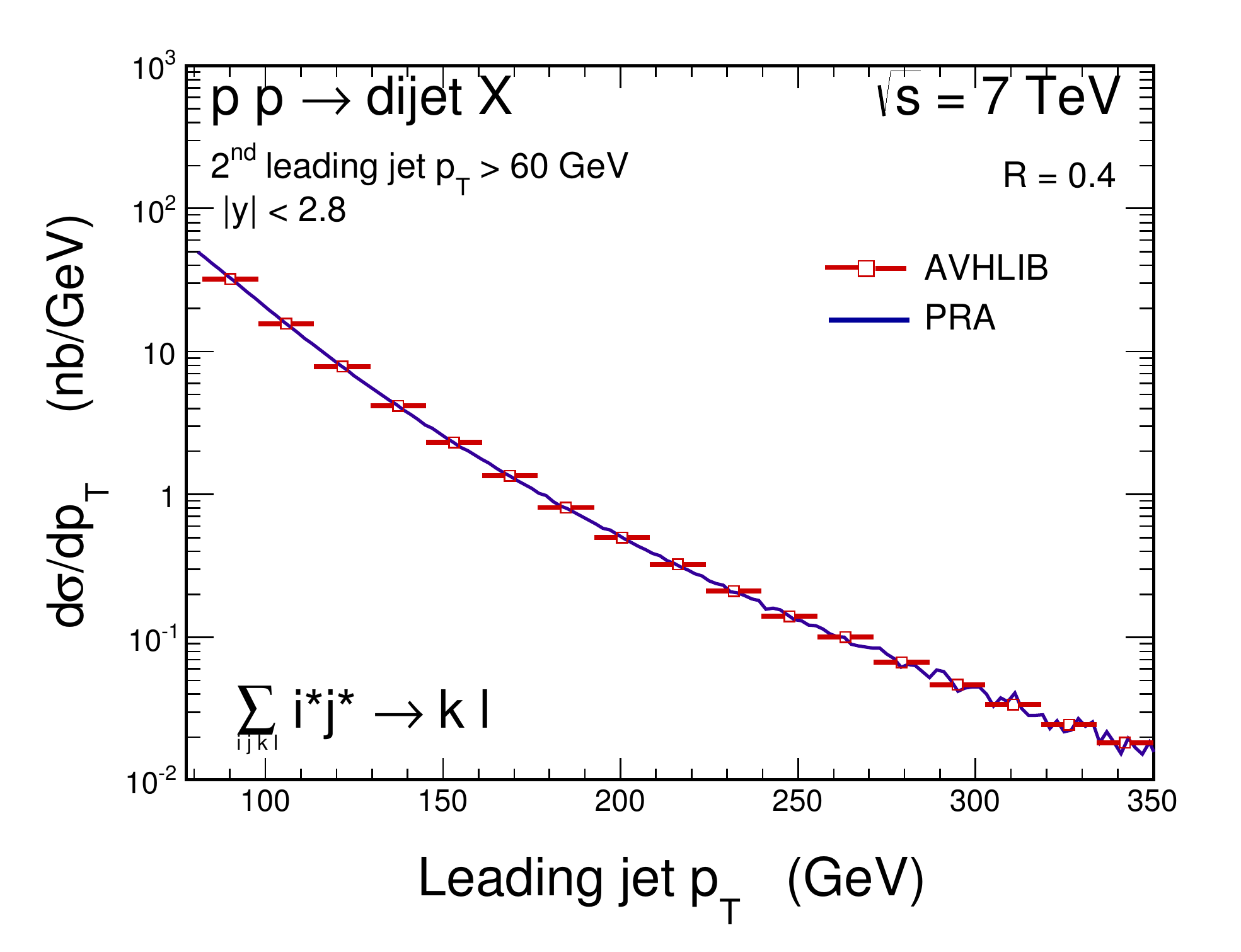}}
\end{minipage}
\hspace{0.5cm}
\begin{minipage}{0.47\textwidth}
 \centerline{\includegraphics[width=1.0\textwidth]{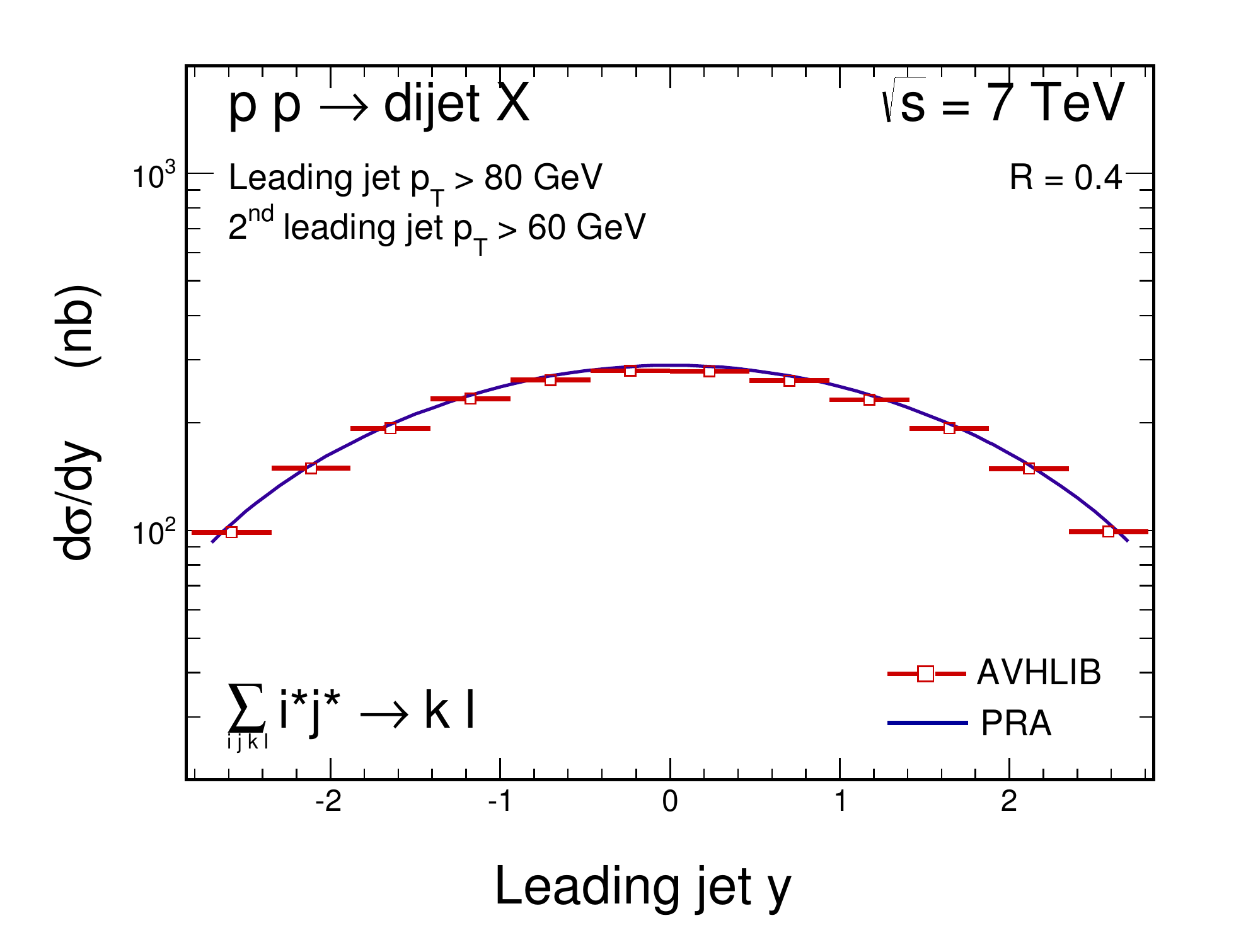}}
\end{minipage}
\caption{
Comparison of the AVHLIB and PRA predictions for dijet production. The leading jet transverse momentum (left panel) and (pseudo)rapidity (right panel) distributions are shown.}
\label{Compar_1}
\end{figure}

\begin{figure}[!h]
\begin{minipage}{0.47\textwidth}
 \centerline{\includegraphics[width=1.0\textwidth]{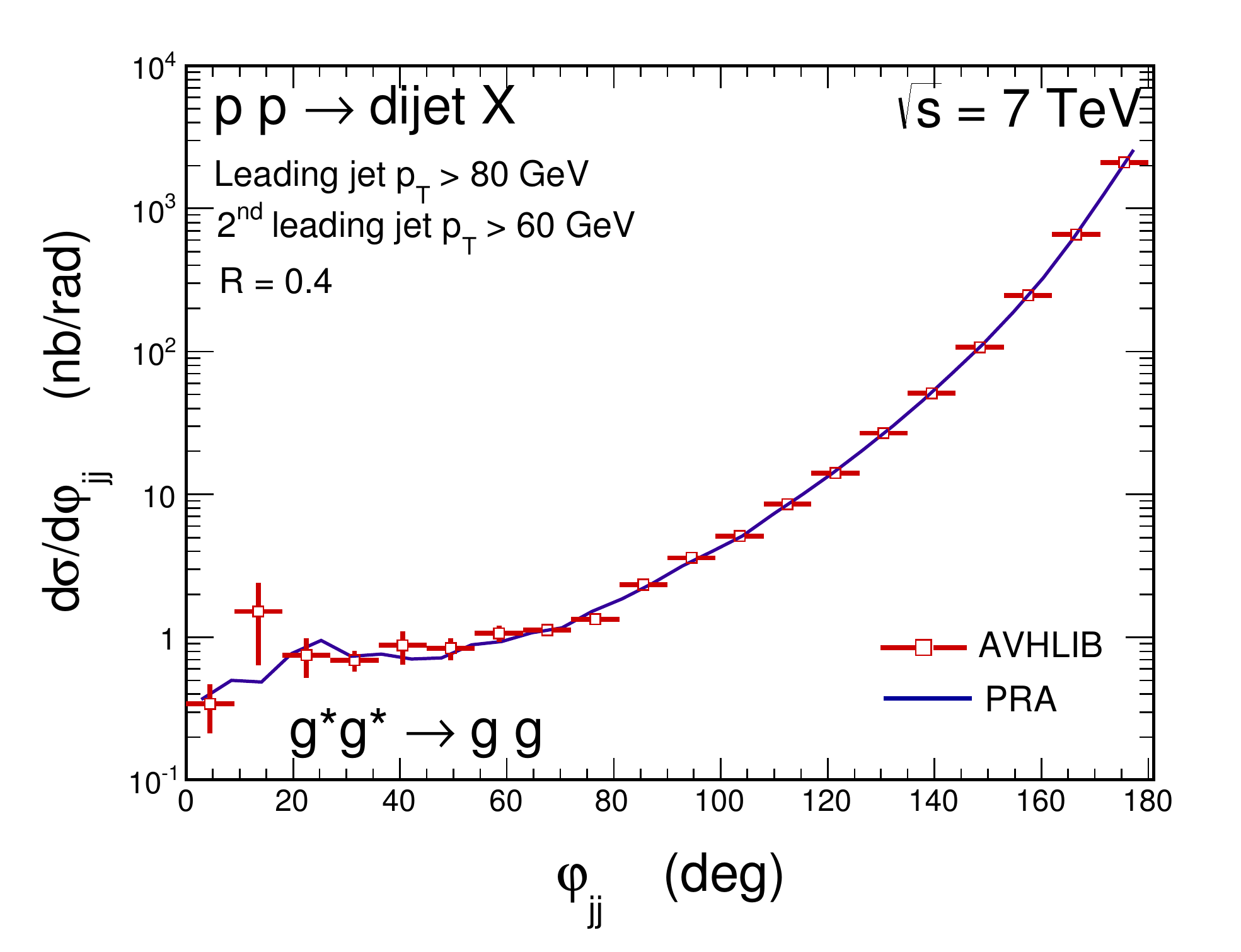}}
\end{minipage}
\hspace{0.5cm}
\begin{minipage}{0.47\textwidth}
 \centerline{\includegraphics[width=1.0\textwidth]{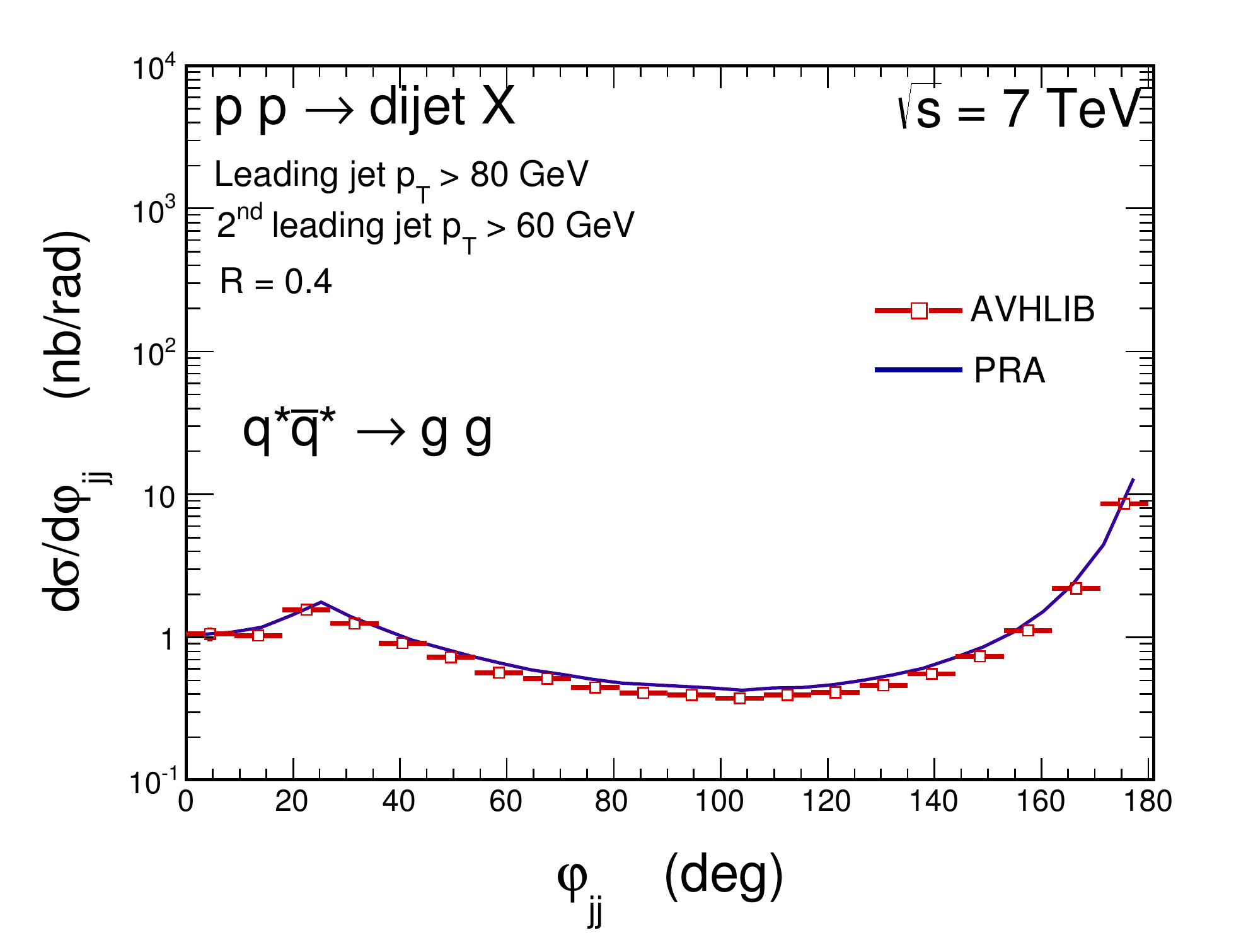}}
\end{minipage}
\caption{
Comparison of the AVHLIB and PRA predictions for azimuthal angle correlation between jets. The calculations for $g^{*}g^{*} \to g g$ (left panel) and $q^{*}\bar{q}^{*} \to g g$ (right panel) subprocesses are shown as an example.}
\label{Compar_2}
\end{figure}

Here we perform a comparison of the cross sections for dijet production obtained both with the QCD amplitudes
from the Parton-Reggeization Approach (PRA) \cite{Nefedov:2013ywa} and those from the AVHLIB, which were in turn cross checked with the results presented in 
Refs.~\cite{vanHameren:2014iua,vanHameren:2015bba}. We find perfect agreement modulo phase space integration uncertainty, 
as shown in Figs. \ref{Compar_1} and \ref{Compar_2}.
The consistency between amplitudes computed in different approaches is, by itself, a non trivial check of the methods employed.
Here we have shown only a few examples. In fact we have checked that agreement is for all possible processes.

\bibliographystyle{MPI2015}
\bibliography{references}

\end{document}